\documentclass[twoside]{article} % arXiv style

\usepackage[bindingoffset=0.in, left=1.in,right=1.in,top=0.75in,bottom=1in, footskip=.25in, headsep=.2in]{geometry} % margins
\usepackage{abstract}
\usepackage{authblk}
\usepackage{blindtext}
\usepackage{fancyhdr}
\usepackage{sectsty}

\sectionfont{\fontsize{12}{15}\selectfont}
\subsectionfont{\fontsize{10}{15}\selectfont}

\newcommand{\s}{\sigma} % sigma

\newcommand{\eq}[1]{\begin{equation}#1\end{equation}}

\usepackage[american]{babel}% hyphenation rules
\addto{\captionsamerican}{}

\usepackage{graphicx}
\usepackage{amsmath}
\usepackage{amssymb} % Mathematical set symbols
\usepackage{upgreek}
\usepackage[normalem]{ulem}
\usepackage{bm} % Bold math
\usepackage[utf8]{inputenc} % Allow for accents
\usepackage[american]{babel} % hyphenation rules
\usepackage[bookmarks=false]{hyperref}
\hypersetup{colorlinks={true},linkcolor={blue},citecolor={blue},urlcolor={blue}}
\usepackage{lscape}

\usepackage{siunitx}

\usepackage{textcomp,gensymb}

\usepackage[square]{natbib}

\usepackage{enumitem}

\def \reff#1 {
	(\ref{#1})
}

\usepackage{soul}
\usepackage[usenames, dvipsnames]{color} %colors

\usepackage{float}
\usepackage{placeins}
\usepackage{caption}

\def \fig [#1][#2]#3#4 {
  \begin{figure}
    \centering
      \includegraphics[width=#2\textwidth]{figs/#1}
      \caption{#3}{\footnotesize{#4}}
      \label{fig:#1}
  \end{figure}
}

\def \subfigs [#1]#2#3#4 {
  \begin{figure}
    \centering
      #2
      \caption{#3}{\footnotesize{#4}}
      \label{fig:#1}
  \end{figure}
}

\def \subfig [#1][#2]#3 {
  \begin{subfigure}[b]{#2\textwidth}
    \includegraphics[width=\textwidth]{figs/#1}
    \caption{#3}
    \label{fig:#1}
  \end{subfigure}
}

\usepackage{etoolbox}

\makeatletter

\patchcmd{\NAT@citex}
  {\@citea\NAT@hyper@{%
     \NAT@nmfmt{\NAT@nm}%
     \hyper@natlinkbreak{\NAT@aysep\NAT@spacechar}{\@citeb\@extra@b@citeb}%
     \NAT@date}}
  {\@citea\NAT@nmfmt{\NAT@nm}%
   \NAT@aysep\NAT@spacechar\NAT@hyper@{\NAT@date}}{}{}

\patchcmd{\NAT@citex}
  {\@citea\NAT@hyper@{%
     \NAT@nmfmt{\NAT@nm}%
     \hyper@natlinkbreak{\NAT@spacechar\NAT@@open\if*#1*\else#1\NAT@spacechar\fi}%
       {\@citeb\@extra@b@citeb}%
     \NAT@date}}
  {\@citea\NAT@nmfmt{\NAT@nm}%
   \NAT@spacechar\NAT@@open\if*#1*\else#1\NAT@spacechar\fi\NAT@hyper@{\NAT@date}}
  {}{}
  
\makeatother

\usepackage{booktabs}

\usepackage{color}
\title{\bf Uncertainty Quantification of Trajectory Clustering Applied to Ocean Ensemble Forecasts}

\author{Guilherme S. Vieira\,$^{1,}$\thanks{salvadorvieira.g@northeastern.edu} }
\author{Irina I. Rypina\,$^{2,}$\thanks{irypina@whoi.edu} }
\author{Michael R. Allshouse\,$^{1,}$\thanks{m.allshouse@northeastern.edu}}
\affil{$^{1}$\,\normalsize{Department of Mechanical and Industrial Engineering, Northeastern University,\\}
\normalsize{Boston, MA 02115, USA\\}
$^{2}$\,\normalsize{Department of Physical Oceanography, Woods Hole Oceanographic Institution,\\}
\normalsize{Woods Hole, MA 02543, USA\\}

\vspace{1em}
Submitted to {\it Fluids} on August 21, 2020
}

\date{}

\begin{document}

\renewcommand{\abstractname}{}    % clear the title
\renewcommand{\absnamepos}{empty} % originally center
 
\maketitle

\vspace{-2em}

%%%%%
%%% ABSTRACT
%%%%%
%%%%%%%%%%%%%%%%%%%%%%%%%%%%%%%%%%%%%%%%%%
% Abstract (Do not insert blank lines, i.e. \\) 
% \clearpage

\begin{abstract}

\begin{center}
    \textbf{Abstract}
\end{center}
\vspace{0.5em}

Partitioning ocean flows into regions dynamically distinct from their surroundings based on material transport can assist search-and-rescue planning by reducing the search domain. The spectral clustering method partitions the domain by identifying fluid particle trajectories that are similar. The partitioning validity depends on the accuracy of the ocean forecasting, which is subject to several sources of uncertainty: model initialization, limited knowledge of the physical processes, boundary conditions, and forcing terms. Instead of a single model output, multiple realizations are produced spanning a range of potential outcomes, and trajectory clustering is used to identify robust features and quantify the uncertainty of the ensemble-averaged results. First, ensemble statistics are used to investigate the cluster sensitivity to the spectral clustering method free-parameters and the forecast parameters for the analytic Bickley jet, a geostrophic flow model. Then, we analyze an operational coastal ocean ensemble forecast and compare the clustering results to drifter trajectories south of Martha's Vineyard. This approach accurately identifies regions of low uncertainty where drifters released within a cluster remain there throughout the window of analysis. Drifters released in regions of high uncertainty tend to either enter neighboring clusters or deviate from all predicted outcomes.

\end{abstract}

% A single paragraph of about 200 words maximum. For research articles, abstracts should give a pertinent overview of the work. We strongly encourage authors to use the following style of structured abstracts, but without headings: (1) Background: Place the question addressed in a broad context and highlight the purpose of the study; (2) Methods: Describe briefly the main methods or treatments applied; (3) Results: Summarize the article's main findings; and (4) Conclusion: Indicate the main conclusions or interpretations. The abstract should be an objective representation of the article, it must not contain results which are not presented and substantiated in the main text and should not exaggerate the main conclusions.

% Keywords
% \keyword{Lagrangian~transport; spectral~clustering; uncertainty quantification; parameter sensitivity; ocean ensemble forecast; drifter~data; search-and-rescue}

% {\it (list three to ten pertinent keywords specific to the article, yet reasonably common within the subject discipline.)}
\section{Introduction}
\label{sec:Intro}

Fluid flows, even if unsteady and aperiodic, may admit persistent patterns generally referred to as coherent structures that reveal flow characteristics related to the transport of fluid particles~\citep{mcwilliams1984emergence,provenzale1999transport,haller2015lagrangian}.
Coherent structures of the elliptic type~\citep{haller12,froyland15,allshouse15} are portions of fluid that do not significantly mix with the rest of the domain.
From a Lagrangian perspective, the perimeter delimiting material within these structures remains nearly constant as they move~\citep{allshouse2012detecting,froyland2015dynamic}, and fluid can be transported over long distances while surrounded by more vigorous mixing~\citep{provenzale1999transport,beron2013objective,haller13}.
Eulerian methods compute coherent structures directly from instantaneous velocity fields and may not highlight features that have a persistent impact on transport.
Additionally, these methods require full velocity fields, which may be unavailable or hard to reconstruct from sparse sets of observed trajectories.
Lagrangian methods, in turn, can identify coherent structures based on the trajectories themselves and identify the dominant features over a given time interval~\citep{provenzale1999transport,boffetta2001detecting,peacock2010introduction,allshouse2012detecting,haller13}.

One approach for identifying these coherent structures uses cluster analysis.
Clustering algorithms reveal underlying structures in data sets by partitioning the data so that similar elements are assigned to the same cluster, while dissimilar elements are assigned to different ones.
These algorithms have been extensively studied and widely applied in image segmentation and anomaly detection, as well as biological and physical processes~\citep{jain1999data,everitt2011cluster}.
For fluid transport analysis, clustering techniques can efficiently identify elliptic structures when applied to fluid particle trajectories~\citep{froyland15,hadjighasem16spectral}, which we refer to as trajectory clustering.

Here, we analyze particle trajectories using the spectral clustering algorithm~\citep{shi2000normalized,ng2002spectral,vonluxburg07,filippone2008survey}.
This method has been used to identify coherent structures in analytic and simulated flows~\citep{hadjighasem16spectral,hadjighasem2017critical,vieira2020internal}.
The clustering performs a systematic partitioning of the trajectories into coherent and incoherent sets, providing a conceptual simplification of the underlying dynamical system for a general flow.
This method is also frame-invariant, and hence the identified structures are the same in all frames that translate or rotate relative to each other.
One drawback of the spectral clustering method, however, is that there are a number of free-parameters that the user has to select, which could impact the results~\citep{hadjighasem2017critical}.

While ocean drifter data can be clustered \emph{a posteriori} to identify coherent structures, an \emph{a priori} analysis to predict coherent structures relies on trajectories obtained from a forecast model.
The clustering accuracy is limited by that of the trajectories, which, in turn, is limited by the accuracy of the velocity model.
In ocean modeling, however, several sources of uncertainties pose a challenge to the Lagrangian approach~\citep{lermusiaux2006uncertainty,lermusiaux2006quantifying}.
To simplify ocean models and reduce computational expenses, the governing equations are only resolved on a restricted range of spatial and temporal scales, and the influence of scales outside this window is either parameterized or neglected.
Uncertainties also arise from the limited knowledge of processes within the scale window, which require approximate representations or parameterizations.
Moreover, measurements used for model initialization and parameter estimation are limited in coverage and accuracy, leading to imprecise initial conditions and model parameters.
Finally, models of interactions between the ocean and the atmosphere are approximate, and ocean boundary conditions are inexact.
All of the above lead to differences between the actual values and the modeled values of physical fields and properties.

To account for model uncertainties intrinsic to the modeling process, an effective option is to perform ensemble statistics.
Different sets of model parameter values generate an ensemble of possible outcomes, which are then processed to provide probabilistic information about the variability on the end results.
Search-and-rescue planning, for instance, already considers ensemble statistics to produce probability-distribution maps for the target’s location~\citep{kratzke2010search}.
The vast uncertain parameter space together with the continuous motion of floating objects driven by unsteady flows, however, can lead to error accumulation in the predicted trajectories~\citep{serra2020search}.
Coherent structures have been shown to depend less on individual trajectories and are more robust to model parameter variations and noise, highlighting the main structures in the flow even in the event of imperfect or scarce trajectories~\citep{haller02,lermusiaux2006quantifying}.
The robustness of the clustering algorithm to perturbations in the individual trajectories via an ensemble set of realizations has yet to be studied and could aid in quantifying uncertainty for the trajectory~clusters.

The detection of robust clusters could then be used in the deployment of drifters to observe such features in the ocean.  
Other Lagrangian methods have been used in the past in experiments to interpret the observed behavior of drifters deployed in the ocean~\citep{olascoaga2013drifter,jacobs2014data,beron2015dissipative,williams2015identifying,rypina2014eulerian,rypina2016investigating,rypina2017trajectory}.
In these studies, drifters were released, tracked, and their trajectories were compared to results from Lagrangian analyses performed \emph{a posteriori} using velocities from models, satellite altimetry or radar measurements.
Few studies, however, have used Lagrangian methods to plan and execute field experiments~\citep{haza2007model,haza2010transport,serra2020search}.
The drifter-release experiment presented in~\citep{filippi2020parameter-free}, whose drifter data is used in this paper, is one of the first experiments targeting coherent structures based on an \emph{a priori} trajectory clustering obtained from forecast simulations.

Our approach accounts for and quantifies uncertainty when clustering trajectories.
We apply the spectral clustering algorithm varying the method free-parameters to understand how the clustering results are sensitive to the implementation, analyze ensemble simulations to understand how the model parameters impact the resulting clusters, and finally apply the method to a forecast data set to compare the clustering prediction to experimental drifter data.
The method by~\citet{hadjighasem16spectral} is modified with a more broadly used similarity function and soft clustering membership probabilities, allowing for a probabilistic view of the clusters for each individual model realization.
Two different systems are analyzed with our approach: an analytic flow model and forecast simulations provided by a coastal ocean model.
First, the Bickley jet analytic flow~\citep{rypina2007lagrangian,beron2010invariant} with model parameter variations is used to mimic model uncertainty. Then, we analyze ensemble-forecasts generated by the MIT-MSEAS ocean model~\citep{haley2010multiscale,haley2015optimizing} of the coastal region of the Martha's Vineyard island.
Ensemble statistics of the resulting clusters provide a probabilistic view of the coherent structures identified by the method.
The forecast clustering results are used to identify coherent structures in a drifter-release experiment, and the drifter trajectories are compared to the forecast cluster behavior and associated uncertainty. 

The paper is organized as follows.
Section~\ref{sec:Method} presents the spectral clustering method used in this work and how clustering results are processed to provide ensemble statistics.
Section~\ref{sec:BJ-Central} introduces the quasi-periodic Bickley jet system, performs ensemble statistics for a single set of particle trajectories while varying the method free-parameters to quantify the sensitivity of the clustering results to the implementation.
Section~\ref{sec:BJ-Ensemble} varies Bickley jet system model parameters to asses the robustness of the identified coherent structures in a scenario of uncertainty in the velocity field.
Section~\ref{sec:MV} presents the clustering analysis used to target coherent structures in a drifter-release experiment south of Martha's Vineyard, and compares the forecast results to the observed drifter trajectories.
Finally, conclusions are presented in Section~\ref{sec:Conclusions}.
\section{Method}
\label{sec:Method}

This section presents the spectral clustering method and how uncertainty is measured.
We adapt the method of~\citet{hadjighasem16spectral} to use soft clustering (fuzzy $c$-means) to assign a cluster membership probability to each particle trajectory.
Section~\ref{subsec:bickley_and_spectral} introduces the method along with specifics of the similarity measure selected and the soft clustering approach.
Section~\ref{subsec:uq_theory} presents the general procedure we use to calculate statistics and quantify uncertainty when considering an ensemble of possible clustering results.

\subsection{Spectral Clustering Method with Soft Memberships for Trajectory Clustering}
\label{subsec:bickley_and_spectral}

To analyze flows from a Lagrangian perspective, we consider massless fluid particles that move with the velocity field $\mathbf{u}(t,\mathbf{x})$.
The trajectory $\mathbf{x}_i(t)$ of fluid particle $i$ departing from $\mathbf{x}_{i}(t_0)$ at time $t=t_0$ is
\eq{
\label{eq:ODE_tracer}
    \mathbf{x}_i(t) = \mathbf{x}_{i}(t_0) + \int_{t_0}^t \mathbf{u}(\tau,\mathbf{x}_i(\tau))d\tau.
}
A total of $N$ particles are initialized in a grid that uniformly covers the targeted domain at $t_0$, and their individual trajectories are numerically integrated and output at discrete time instances within the time interval $[t_0,\,t_f]$. 

We use the set of trajectories $\{\mathbf{x}_i(t)\}_{1\leq i\leq N}$ to partition the spatial domain into clusters.
The spectral clustering algorithm performs an eigenanalysis to project the trajectory set onto a subspace that may yield clusters maximizing the within-cluster similarity and minimizing the between-clusters similarity~\citep{nock2009soft}.
Particles clustered together should move as a compact group, with limited mixing with particles outside of the cluster, while particles in different clusters should experience dissimilar trajectories~\citep{froyland15, hadjighasem16spectral}.

The spectral analysis requires the construction of a positive, weighted, undirected graph known as the similarity graph.
This graph quantifies the pairwise similarities between trajectories.
Each graph node represents a particle trajectory, and the graph edges are weighted by the similarity between the nodes they connect~\citep{ vonluxburg07,hadjighasem16spectral,shi2000normalized}. 
To compute these weights, one must first define a measure for similarity between trajectories. One possible metric for the dissimilarity between trajectories $\mathbf{x}_i$ and $\mathbf{x}_j$ is their time-averaged distance
\eq{
r_{ij} = \frac{1}{t_f-t_0}\int_{t_0}^{t_f}\textrm{dist}(\mathbf{x}_i(\tau), \mathbf{x}_j(\tau)) d\tau,
\label{eq:rij}
}
where $\textrm{dist}(\cdot,\cdot)$ is the Euclidean distance.
Then, a similarity measure for the edge weights $w_{ij}=w(r_{ij})$ must be chosen as a function of the time-averaged distance.
The only restriction on this functional dependence is that the weight should be a monotonically decreasing function of the distance.
This choice of similarity function controls the graph edge weight distribution, and therefore can impact the clustering results~\citep{vonluxburg07}.

\citet{hadjighasem16spectral} use a similarity function that is the inverse of the time-averaged distance,  $w_{ij}=l_x/r_{ij}$ (a constant $l_x$ is included here to make the weight dimensionless and does not impact the results). 
A more widely used function for graph partitioning is the Gaussian function~\citep{shi2000normalized,ng2002spectral,vonluxburg07}
\eq{
\label{eq:Gaussian}
w_{ij} = \exp\left(- r^2_{ij}/2\sigma^2\right),
}
which we will use.
The similarity radius $\s>0$ is a method free-parameter that controls the spatial width of the connected neighborhoods~\citep{shi2000normalized}.
Note that $w_{ii}=1$ and $w_{ij}\to0$ for $r_{ij}\gg\s$.
The choice of the Gaussian measure has a number of advantages over the $l_x/r_{ij}$ choice: first, it is bounded, with $0\leq w_{ij}\leq 1$; second, no offset value is set in the diagonal entries~\citep{hadjighasem16spectral}; and third, the sparsification step can be skipped (or at least has a negligible impact on the results, see Supplementary Material A), as \eqref{eq:Gaussian} goes to zero faster than the measure in~\citep{hadjighasem16spectral} for large $r_{ij}$.
The similarity radius $\s$ in \eqref{eq:Gaussian} determines the rate of decay of $w_{ij}$ to zero as $r_{ij}$ increases, which is analogous to the graph sparsification in~\citep{hadjighasem16spectral} where $w_{ij}=0$ for $r_{ij}$ above a defined threshold.
A direct comparison between these two similarity functions and the impact of the $\s$-choice are presented in Sections~\ref{subsec:parameter_gaussian} and \ref{subsec:parameter_sigma}, respectively.

The similarity matrix $\mathbf{W}~\in~\mathbb{R}^{N\times N}$ stores the $w_{ij}$ values, and the diagonal degree matrix $\mathbf{D}$ is computed such that $d_{ii}=\sum_{j=1}^{N}w_{ij}$.
The generalized eigenvectors $\mathbf{q}$ of the unnormalized graph Laplacian $\mathbf{L} = \mathbf{D} - \mathbf{W}$ are computed from the generalized eigenproblem
\eq{
\label{eq:laplacian_eigenproblem}
\textbf{L}\mathbf{q} = \lambda \textbf{D} \mathbf{q}.
}
The normalized vectors $\mathbf{q}_1,\mathbf{q}_2,\ldots,\mathbf{q}_N$, corresponding to the generalized eigenvalues ${0\leq\lambda_1\leq\ldots\leq\lambda_N}$, differentiate properties in the graph and facilitate the clustering process~\citep{vonluxburg07}.  
While all of the eigenvectors provide information, the dominant eigenvectors, associated with the smallest eigenvalues, reveal the most dynamically relevant characteristics.
Each trajectory is characterized by a value within each eigenvector, and these are the values ultimately used to cluster~trajectories.  

The eigenvector matrix $\mathbf{Q}~\in~\mathbb{R}^{N\times M}$ whose columns are the dominant eigenvectors $\mathbf{q}_1,\ldots,\mathbf{q}_M$, with $M\ll N$, stores the characteristics used to cluster the trajectories.
The number of retained eigenvectors $M$ is specified depending on the system.
Let $\mathbf{y}_i \in \mathbb{R}^{M}$ be the characterization vector corresponding to the $i$-th row of $\mathbf{Q}$, which contains the condensed differentiating information of trajectory $i$.
The eigenanalysis provides a suitable low dimensional representation of the data set.
The characterization vectors $\{\mathbf{y}_i\}_{1\leq i \leq N}$ are then partitioned into $K$ clusters.
The relationship between the number of clusters $K$ and the number of dominant eigenvectors $M$ used for clustering depends on characteristics of the system, and is specified for each of the case studies in the following sections.

We cluster the characterization vectors using a fuzzy \textit{c}-means algorithm~\citep{froyland15,filippone2008survey}, instead of the conventional $k$-means often performed at this step~\citep{vonluxburg07,hadjighasem16spectral}.
The $c$-means algorithm assigns to each trajectory $i$ probabilities $p_{i,k}\in[0,1]$ of being a member of cluster $k$, with $1\leq k\leq K$, such that $\sum_{k=1}^{K}p_{i,k}=1$. 
This step requires the prescription of a fuzziness parameter $m>1$ that controls how ``tight'' the clustering membership probabilities are.  
As $m\to1$, the clustering result approaches the \textit{k}-means result where $p_{i,k}\in\{0,1\}$.
For greater $m$, the ``looser'' distribution in membership probabilities allows the identification of particles that present intermediate behaviors between clusters, as opposed to always assigning a trajectory as member of a single cluster.
The cluster centers are initialized based on the dominant eigenvectors before starting the \textit{c}-means iterative process detailed by~\citet{froyland15}.
Finally, \citet{hadjighasem16spectral} use an eigengap heuristic to determine $M$ and $K$ that relies on a large cluster differentiation to work properly, and in many practical cases an eigengap may be less pronounced or nonexistent~\citep{vonluxburg07}. 
% We choose to leave the number of clusters $K$ as another method free-parameter 
We leave the number of dominant eigenvectors and clusters as method free-parameters and study the related uncertainty associated to the clustering results for each $K$-choice.

\subsection{Uncertainty Quantification for Multiple Realizations}
\label{subsec:uq_theory}

In Sections~\ref{sec:BJ-Ensemble} and \ref{sec:MV}, we will be interested not in applying the spectral clustering algorithm to a single realization of the flow, but in collecting clustering results from different realizations and combining them to get statistical information about the variability of the clusters with the method free-parameters $\s$, $m$ and $K$, or with velocity model parameters.
We therefore need a method to combine clustering results from different realizations.
Regardless of the differences between realizations, we initialize the $N$ particles on the same grid, so that trajectory $i$ is uniquely labelled across realizations, based on $\mathbf{x}_i(t_0)$.
Each realization $I\in\{1,\ldots,R\}$ generates a full set of membership probabilities $p_{i,k}^{(I)}$, with $i\in\{1,\ldots,N\}$ and $k\in\{1,\ldots,K\}$.
The cluster labels for different realizations are matched based on the similarity between cluster centers.

To quantify the uncertainty of a trajectory $i$ being a member of cluster $k$, the probabilities $p_{i,k}^{(I)}$ are used to compute the mean membership probabilities over all $R$ realizations
\eq{
\overline p_{i,k} = \frac1{R} \left( \sum_{I=1}^{R} p^{(I)}_{i,k} \right),
\label{eq:mean}
}
and the corresponding sample standard deviations
\eq{
S_{i,k} = \sqrt{\frac{1}{R-1} \sum_{I=1}^{R} \left(p^{(I)}_{i,k} - \overline p_{i,k}\right)^2}.
\label{eq:std}
}
Both the mean and the standard deviation of the realizations are bounded values, with $0\leq \overline p_{i,k} \leq 1$ and $0\leq S_{i,k} \leq 0.5$.
We perform this calculation for each trajectory, for each cluster.

\section{The Bickley Jet System and Sensitivity to Method Free-Parameters}
\label{sec:BJ-Central}

We analyze the analytic, quasi-periodic Bickley jet system to evaluate the spectral clustering method free-parameter and velocity field model sensitivity.
This system is an idealized model of a meandering zonal jet under geostrophic balance and has been extensively used to illustrate coherent structures in fluid flows~\citep{rypina2007lagrangian,beron2010invariant,schlueter2017coherent,hadjighasem2017critical}.
The model features a sheared zonal flow on which a superposition of Rossby-like waves propagate.
The streamfunction $\psi$ prescribing the two-dimensional velocity field $\mathbf{u}=-\partial_y \psi \,\mathbf{e}_x + \partial_x\psi \,\mathbf{e}_y$ with a superposition of three waves is
\eq{
\psi(t,x,y) =  -UL\tanh \left(\frac{y}{L}\right) + UL\,\mathrm{sech}^2 \left(\frac{y}{L}\right) \sum_{n=1}^{3}A_n \cos\left[k_n(x-c_nt+\phi_n)\right],
}
where $U$ and $L$ are a characteristic speed and length, respectively.
The domain is periodic in the $x$-direction, with periodicity $l_x=2\pi R_e\cos(60^\circ)$, where $R_e=\SI{6371}{km}$ is Earth's radius. 
The rectangular domain corresponds to $x/l_x\in[0,1]$, and we limit our analysis to $y/l_x\in[-0.15, 0.15]$. 
The Rossby-like waves correspond to the three longest wave modes in the periodic domain, with amplitudes $A_n$, wave numbers $k_n=2\pi n/l_x$, phase speeds $c_n$, and phases $\phi_n$, for $n=1,2,3$.
To model the self-consistent state obtained by~\citet{rypina2007lagrangian} for modes 2 and 3 on the periodic domain, we fix $U=\SI{62.74}{m s^{-1}}$, $L=\SI{1767}{km}$, $c_2/U=0.2051$ and $c_3/U=0.4615$. The mode-1 wave has speed $c_1/U= 0.1446$ chosen based on the golden ratio to break periodicity~\citep{rypina2007lagrangian,hadjighasem16spectral}.
Provided that $A_1\ll \min(A_2,A_3)$, the mode-1 wave acts as a small perturbation to the system's periodicity.
In this section, we fix the mode amplitudes to $A_1 = 0.0075$, $A_2 = 0.15$, and $A_3 = 0.30$, values used by~\citet{hadjighasem16spectral}, and all three waves are in phase: $\phi_1=\phi_2=\phi_3=0$. 
Note, however, that while the system dynamics depend on the values of $A_n$ and $\phi_n$~\citep{rypina2007lagrangian}, there is no physical basis for the stated choice of amplitudes and phases, and the impact of varying these parameters will be explored in Section~\ref{sec:BJ-Ensemble}.

The present section discusses the application of the spectral clustering algorithm to the Bickley jet system, and studies the sensitivity of the results to user-defined method free-parameters.
We present the impact of the choice of similarity measure in Section~\ref{subsec:parameter_gaussian}.
Then, study the sensitivity to the method free-parameters: similarity radius $\s$, in Section~\ref{subsec:parameter_sigma}, tightness of the cluster memberships $m$, in Section~\ref{subsec:parameter_m}, and number of clusters $K$, in Section~\ref{subsec:parameter_K}.

\subsection{Gaussian Similarity Measure}
\label{subsec:parameter_gaussian}

\begin{figure}[t]
\centering
\includegraphics[width=\textwidth]{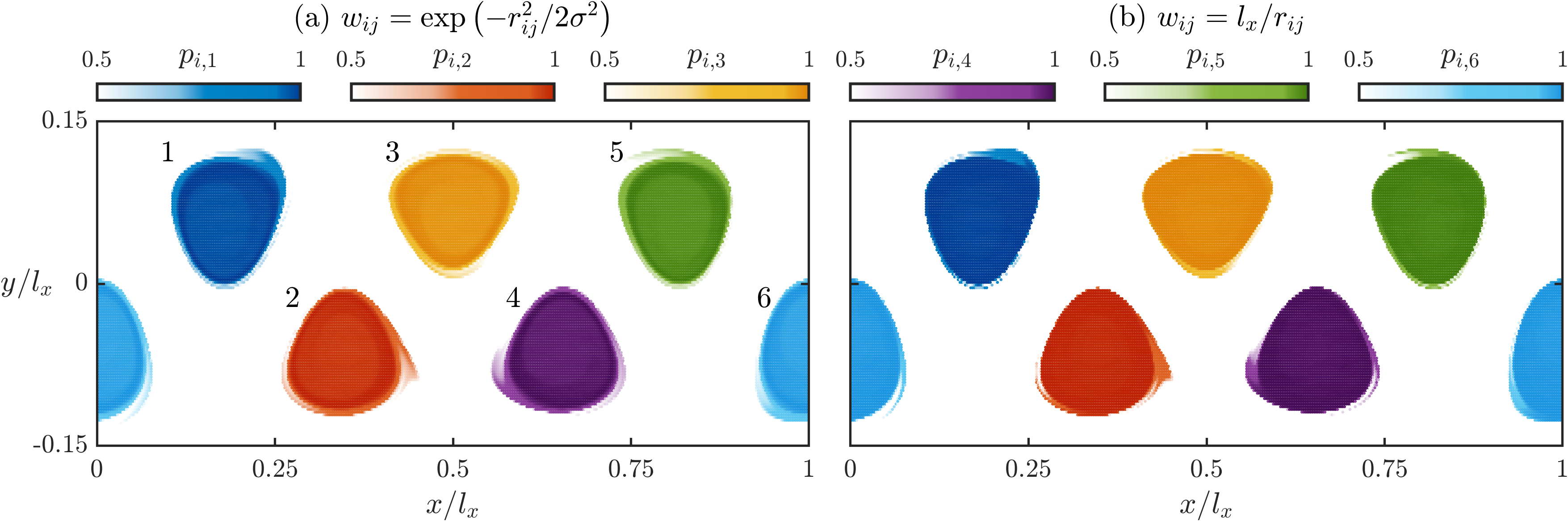}
\caption{Spectral clustering membership probabilities for clusters $k \in \{1,\ldots,6\}$ identifying materially coherent vortices, with fuzziness $m=2$ and number of clusters $K=7$ (incoherent cluster probabilities are omitted). The similarity functions $w_{ij}$ used are (\textbf{a}) the proposed Gaussian similarity measure \eqref{eq:Gaussian} with $\s/l_x=0.020$, and (\textbf{b}) the inverse measure $l_x/r_{ij}$ from~\citep{hadjighasem16spectral} sparsifying values less than 1/0.075.
The six color maps presented here are also used in Figures \ref{fig2:sigma_representatives}, \ref{fig4:}(a,b) and \ref{fig6:diff_clusters}.
}
\label{fig1:hadjighassem}
\end{figure}  

To cluster the Bickley jet system according to the method described in Section~\ref{subsec:bickley_and_spectral}, particles are initialized in a uniform grid of 400 by 120 positions uniformly covering the domain, and advected from $t_0=0$ to $t_f=40\,$days, matching~\citep{hadjighasem16spectral}.
The distance function in \eqref{eq:rij} takes into consideration the $x$-periodicity of the domain, and the $M=6$ dominant eigenvectors of \eqref{eq:laplacian_eigenproblem} are used to partition the domain into $K=7$ clusters, to account for 6 materially coherent vortices, which are the coherent clusters, and an incoherent cluster, the chaotic sea~\citep{hadjighasem16spectral}.
The method uses a fuzziness parameter $m=2$.

The membership probabilities for the clusters identifying the 6 materially coherent vortices at time $t_0$ are presented in Figure~\ref{fig1:hadjighassem}.
Figure~\ref{fig1:hadjighassem}(a) presents our results, obtained using a Gaussian similarity function \eqref{eq:Gaussian}, with similarity radius $\sigma/l_x=0.020$, and no sparsification of the similarity matrix.
The membership probabilities, plotted in different colors, highlight the 6 coherent vortices.
We assign labels to the vortices, from left to right, and vortex 6 appears split at $t_0$ due to the $x$-periodicity.
The membership probabilities of being part of the chaotic sea, the seventh cluster, are complementary to the ones plotted and are omitted throughout.
Note that the use of a soft membership assigns to particles located at the periphery of the vortices lower probabilities of being a member, which relates to lower similarity in the dynamics (some of them may, for instance, be trapped inside the vortices for just a fraction of the time window of analysis, then merge with the chaotic sea, or vice-versa).

The results for the Gaussian similarity measure in Figure \ref{fig1:hadjighassem}(a) are similar to those obtained with the $l_x/r_{ij}$ similarity measure used by~\citet{hadjighasem16spectral}, presented in Figure \ref{fig1:hadjighassem}(b).
For the latter, matrix entries $w_{ij}=0$ for $r_{ij}/l_x>0.075$, and the offset diagonal value is chosen as 100 times the largest matrix entry.
The clustering results are particularly sensitive to the sparsification threshold, which relates to the choice of $\s$ for the similarity measure \eqref{eq:Gaussian}.
Both similarity functions yield similar results for the selected parameter values, with smoother transitions in membership probabilities (from 1 to 0) for the Gaussian measure.

When using the $l_x/r_{ij}$ similarity measure, the degree of sparsification is an additional parameter for the clustering method that can impact the results.  
This parameter can be eliminated for the Gaussian similarity measure as no sparsification was necessary for the result in Figure \ref{fig1:hadjighassem}(a), but it is worthwhile to sparsify the matrix to reduce computational costs and storage.
We demonstrate in the Supplementary Material A that sparsifying entries of $\mathbf{W}$ that satisfy ${w_{ij}<\exp(-4^2/2)\approx\SI{3e-4}{}}$, corresponding to $r_{ij} > 4\sigma$, has negligible impact on the clustering membership probability results, and hence we sparsify according to this rule hereafter.
Note that this result is valid for the Bickley jet, but the impact of sparsification may vary for an arbitrary flow.

\begin{figure}
\centering
\includegraphics[width=\textwidth]{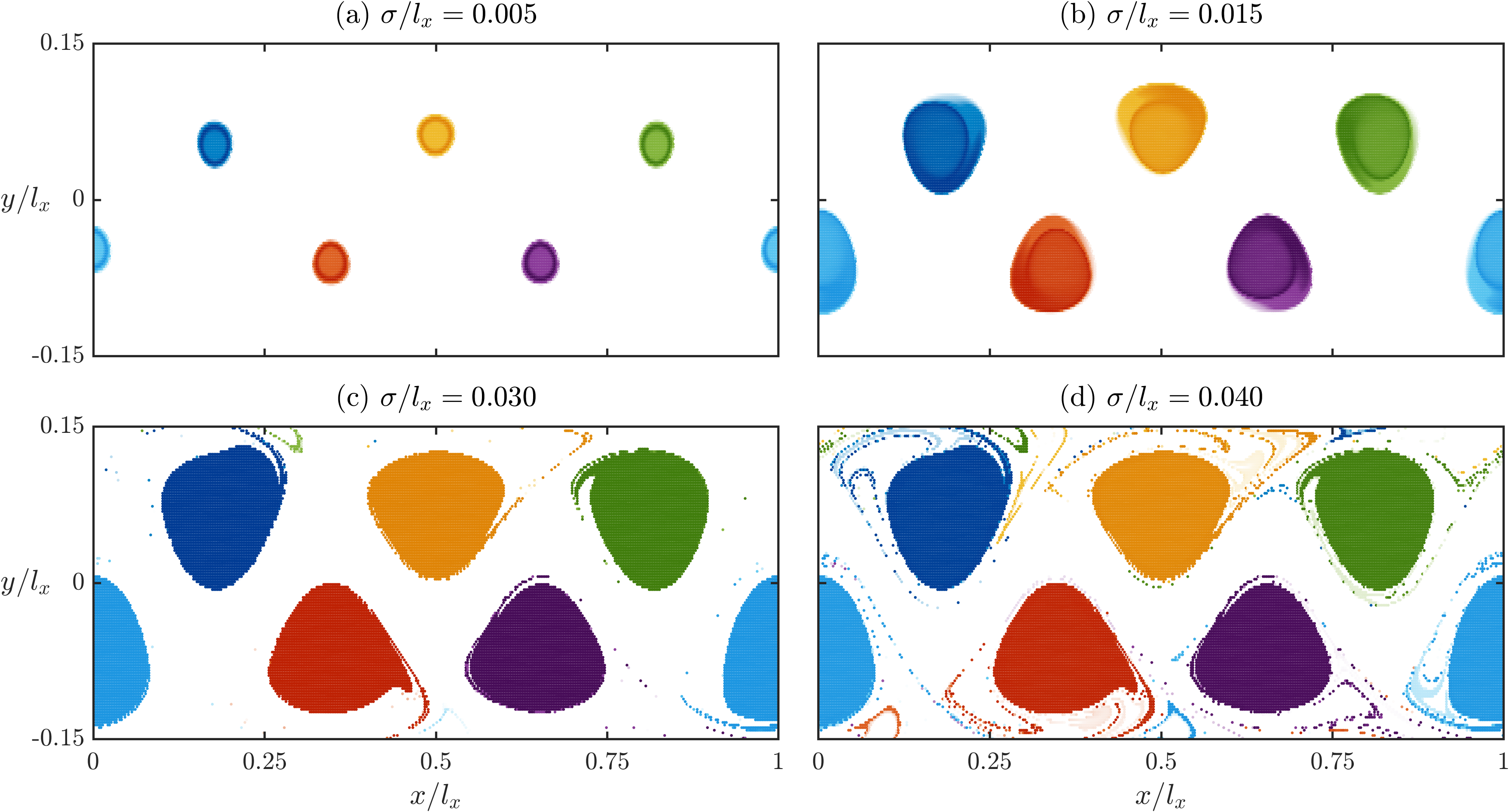}
\caption{Membership probabilities for vortices $k \in \{1,\ldots,6\}$, for different choices of the parameter $\s$, plotted at $t=t_0$. Values correspond to
(\textbf{a}) $\sigma/l_x=0.005$, (\textbf{b}) $0.015$, (\textbf{c}) $0.030$ and (\textbf{d}) $0.040$. Corresponding colorbars presented in Figure~\ref{fig1:hadjighassem}. (See Supplementary Material for a video of the time evolution of the clustered trajectories.)}
\label{fig2:sigma_representatives}
\end{figure}

\subsection{Similarity Radius}
\label{subsec:parameter_sigma}

We highlighted in the previous section how the similarity radius $\s$ is closely related to the sparsification threshold used by~\citet{hadjighasem16spectral}.
While the sparsification threshold selection was mentioned in~\citep{hadjighasem16spectral}, the clustering results are highly sensitive to this parameter~\citep{filippi2020parameter-free}, and we now address this sensitivity.

While there may be some intuition on the size of the structures of interest, this is not always helpful in prescribing $\s$ or in understanding how the $\s$-choice impacts the resulting clusters.
\citet{hadjighasem16spectral} choose their threshold by defining which values of $\mathbf{W}$ to keep based on a fixed percent sparsification of the matrix.
However, for a fixed percent sparsification, the graph connections that are retained are ultimately a function of the number of particles and their distribution in the initial grid.

Based of this relationship between the sparsification level in~\citep{hadjighasem16spectral} and the parameter $\s$ in the Gaussian similarity function \eqref{eq:Gaussian}, we vary $\s$ to demonstrating the clustering sensitivity to changes in sparsification.
First, we define an interval bounding all relevant $\sigma$-values to be tested. 
For $\sigma/l_x$ distributed on the interval $[0.005,0.040]$ with steps of 0.001 (36 cases), the method is applied with $m=2$ to identify $K=7$ clusters.
Figure \ref{fig2:sigma_representatives} presents the membership probabilities for each of the six coherent clusters for $\sigma/l_x=0.005$, 0.015, 0.030 and 0.040.
Compared to the results for $\sigma/l_x=0.020$ presented in Figure \ref{fig1:hadjighassem}(a), smaller values of $\s$ (Figures~\ref{fig2:sigma_representatives}(a,b)) tend to shrink the coherent clusters to the vortex cores only, while larger choices of $\s$ (Figures~\ref{fig2:sigma_representatives}(c,d)) assign higher membership probabilities to filaments that correspond to particles that do not belong to the coherent vortex from the start, but have a long residence time on the perimeter of the vortex.

\begin{figure}
\centering
\includegraphics[width=\textwidth]{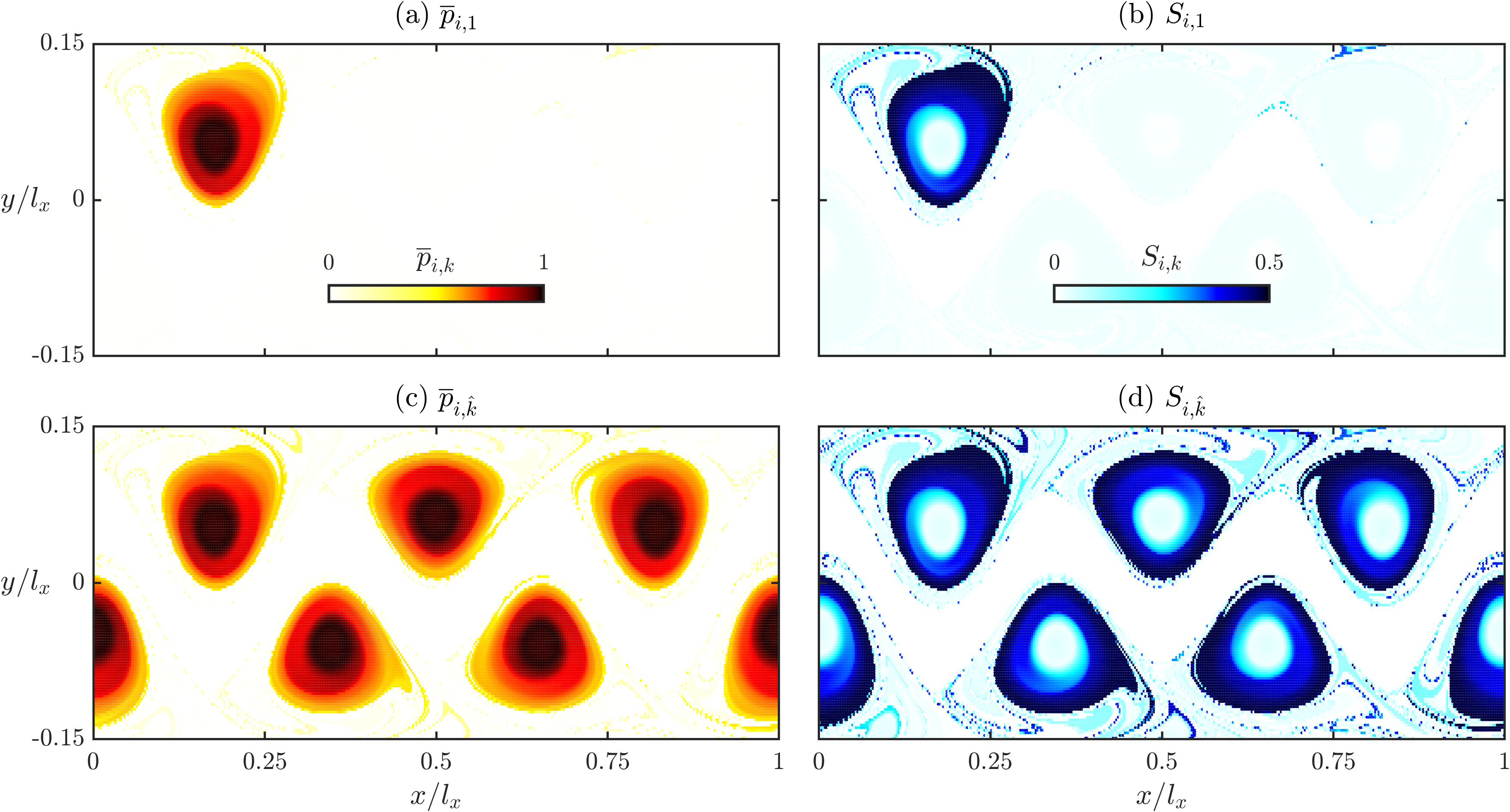}
\caption{Clustering statistics varying $\s$, plotted at $t=t_0$, for $\sigma$ uniformly distributed in $[0.005,0.040]$. Membership probability (\textbf{a}) mean for $k=1$, (\textbf{b}) standard deviation for $k=1$, (\textbf{c}) superimposed mean and (\textbf{d}) superimposed standard deviation for $k \in \{1,\ldots,6\}$. The two color maps presented here are also used in Figures \ref{fig4:}(c,d), \ref{fig5:}, \ref{fig7:} and \ref{fig8:demonstration}(a).
}
\label{fig3:}
\end{figure} 

To determine how the membership probabilities for each trajectory depend on $\s$, we use the information from the different realizations to compute the membership probability means $\overline p_{i,k}$ and sample standard deviations $\overline S_{i,k}$ for each trajectory and cluster, as described in Section~\ref{subsec:uq_theory}.
These statistical measures are presented in Figure \ref{fig3:}.
In Figure \ref{fig3:}(a), we present $\overline p_{i,1}$ for vortex $k=1$, and in Figure \ref{fig3:}(b) the corresponding standard deviation $S_{i,1}$.
Figures \ref{fig3:}(c,d) condense the information for all vortices $k \in \{1,\ldots,6\}$.
The superimposed mean $p_{i,\hat{k}}$ in Figure \ref{fig3:}(c) and the  superimposed standard deviation $S_{i,\hat{k}}$ in Figure \ref{fig3:}(d) are such that, for each trajectory $i$, $\hat{k}$ is the cluster that maximizes the mean membership probability, hence $\hat{k}=\textrm{argmax}_{k}\,\overline p_{i,k}$.

Figure \ref{fig3:}(a,c) demonstrates that the vortex cores have the greatest mean membership, which relates to the fact that those particles are identified with high membership probabilities in all realizations.
Particles further away from the cores have a lower mean, as a result of lower probabilities of being part of the respective vortices, in particular for low $\sigma$ values.
We also notice sharp drops in the probability after a certain vortex size.
The uncertainty of particles being part of a vortex is highlighted by the standard deviations in Figure \ref{fig3:}(b,d), where we again see negligible standard deviation (low uncertainty) on the membership probabilities for the cores.
Because there are no realizations that identify the central jet and some portions of the chaotic sea as part of the clusters, there is also a low standard deviation for those trajectories.
The highest uncertainty is obtained for particles between the core and the chaotic sea, and some filaments are also highlighted with higher standard deviation.
Those filaments correspond to particles consistently identified as part of a vortex, with high membership probabilities, for $\s$ greater than a threshold, but not for lower values of $\s$.

\subsection{Fuzziness Parameter}
\label{subsec:parameter_m}

\begin{figure}[t]
\centering
\includegraphics[width=\textwidth]{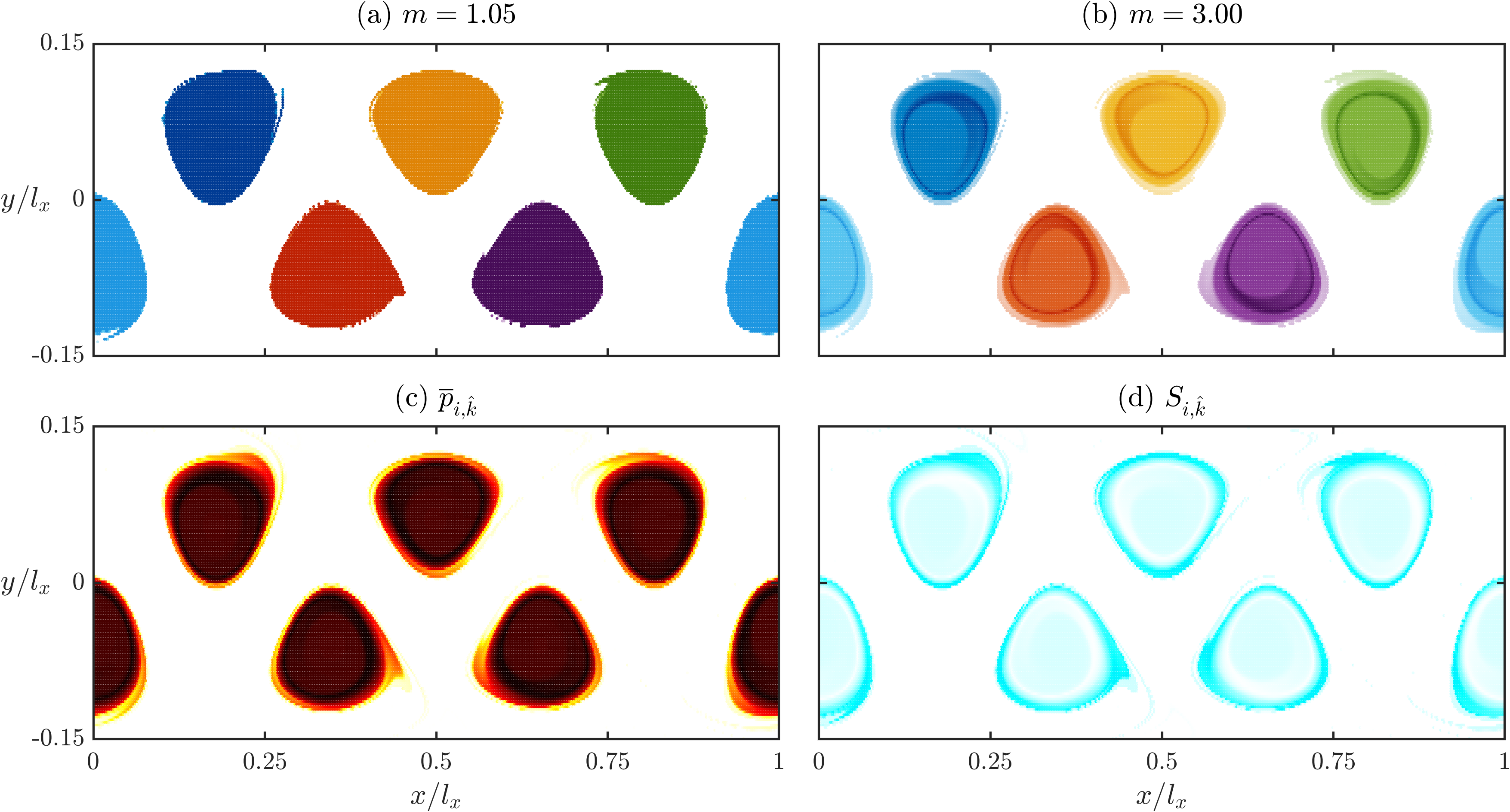}
\caption{Membership probabilities for $k \in \{1,\ldots,6\}$, with (\textbf{a}) $m=1.05$, and (\textbf{b}) $m=3.00$. Clustering statistics plotted at $t=t_0$ for $m$ uniformly distributed in $[1.05,3.00]$. Membership probability (\textbf{c}) superimposed mean and (\textbf{d}) superimposed standard deviation for $k \in \{1,\ldots,6\}$. Corresponding colorbars presented in Figures~\ref{fig1:hadjighassem} and \ref{fig3:}.}
\label{fig4:}
\end{figure} 

Next, we consider the sensitivity of the choice of the fuzziness parameter $m$, introduced by the $c$-means step that assigns cluster membership probabilities.
In this analysis, $\s/l_x=0.020$ while $m$ is varied on the interval $[1.05,3.00]$ with steps of 0.05.
The results are presented in Figure \ref{fig4:}.
The membership probabilities for the extreme values ($m=1.05$ and 3.00) are illustrated in Figure \ref{fig4:}(a,b), and the superimposed mean and standard deviation in Figure \ref{fig4:}(c,d).

While $m=1.05$ (in Figure \ref{fig4:}(a)) yields a bimodal probability distribution, with $99\%$ of the $p_{i,k}>0.5$ cases also being greater than $0.95$ (so tending to the corresponding $k$-means result), the use of $m=3.00$ (Figure \ref{fig4:}(b)) produces smoother probability transitions from the vortex core to the perimeter, as well as overall lower membership probabilities, with only $10\%$ of the $p_{i,k}>0.5$ cases greater than $0.95$.
The superimposed mean and standard deviations reveal a similar trend to the one observed for the dependence on $\s$, with two major differences: \emph{(i)} while varying $m$ introduces uncertainty in the membership probability of trajectories starting at the vortex perimeters, the $m$-choice does not lead to the same significant capture of filaments as member of the vortex for the current $\s/l_x=0.020$ value, and \emph{(ii)} the magnitude of the standard deviations related to $m$ are about half of the ones associated to $\s$ (see Figure \ref{fig4:}(d) in comparison to Figure \ref{fig3:}(d)).

\subsection{Number of Clusters}
\label{subsec:parameter_K}

Finally, the number of clusters in the system is not necessarily known beforehand, and here we address how the clustering results statistically vary for different choices of $K$.
As presented in~\citep{hadjighasem16spectral}, for a suitable sparsification level of the similarity matrix, the eigengap heuristic can be used to infer the use of $M=6$ dominant eigenvectors to choose $K=7$ for the Bickley jet, and thus identify the 6 materially coherent vortices, plus the chaotic sea.
It is known, however, that the existence of an eigengap is not guaranteed for any system, and depends on, among other properties, the connectivity of the similarity graph~\citep{vonluxburg07}.
The number of clusters for systems not as distinctly separated as the Bickley jet will, therefore, be more challenging to determine.
Without any prior knowledge about the number of clusters in the system, one could consider clustering based on different numbers of dominant eigenvectors of \eqref{eq:laplacian_eigenproblem} into different $K$, which might result in merging clusters, splitting clusters, missing clusters, or even identifying new ones.
As we vary $K$, we fix the relationship between $M$ and $K$ to $K=M+1$ to always include the chaotic sea as an independent cluster~\citep{hadjighasem16spectral}.

Because the free-parameter $K$ cannot be varied continuously, and because varying $K$ while fixing $\s$ and $m$ would only mean changing the number of eigenvectors to use in the $c$-means step, we adopt a different strategy to quantify the $K$-uncertainty.
For each choice of $K$, we perform statistics using realizations in which $\s$ and $m$ are varied. 
The $(\s/l_x,\,m)$ pairs are drawn from uniform distributions over $[0.005,0.040]\times [1.05,3.00]$,  corresponding to the same intervals in Sections \ref{subsec:parameter_sigma} and \ref{subsec:parameter_m}.
We consider 100 samples drawn from uniform distributions, and compute mean and sample standard deviations for $K=6$, 7 and 8.
One should now expect missing or merging vortices for $K<7$, while splitting vortices or identifying new structures for $K>7$.
For the ensemble statistics, if vortex $k$ is not identified in realization $I$, we set $p^{(I)}_{i,k}=0$ for all $i$.
For $K=8$, a new coherent cluster corresponding to the jet is consistently identified, in all realizations. The jet is therefore considered a seventh coherent cluster.

\begin{figure}[t]
\centering
\includegraphics[width=\textwidth]{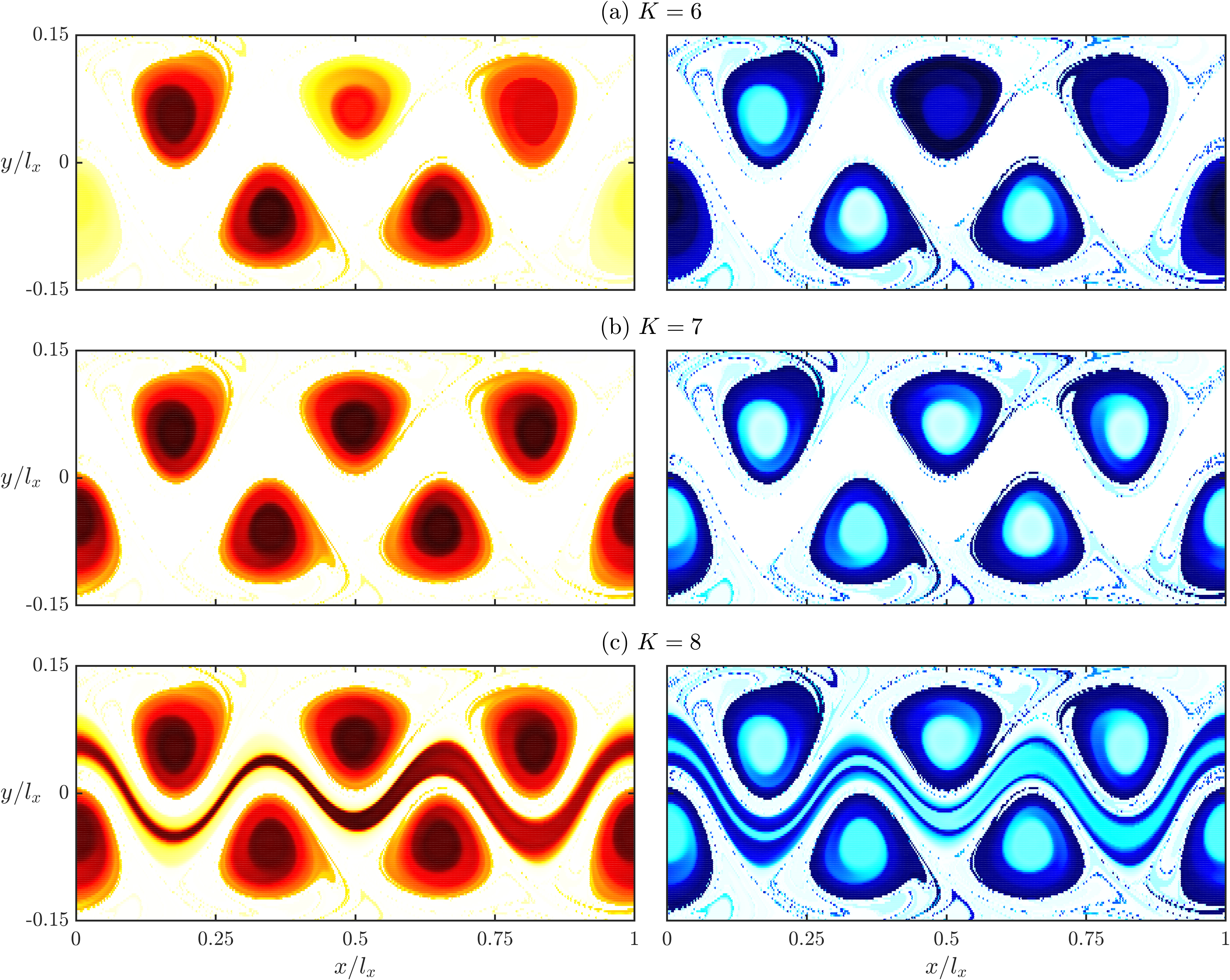}
\caption{Clustering statistics varying $\s$ and $m$, for different numbers of clusters $K$, with  $(\s/l_x,\,m)$ drawn from a uniform distribution over $[0.005,0.040]\times [1.05,3.00]$. Each row presents the superimposed mean $\overline{p}_{i,\hat{k}}$ (left) and the corresponding superimposed standard deviation $S_{i,\hat{k}}$ (right), for (\textbf{a})~$K=6$, (\textbf{b})~$K=7$ and (\textbf{c})~$K=8$. Corresponding colorbars presented in Figure~\ref{fig3:}.}
\label{fig5:}
\end{figure} 

The superimposed means and standard deviations for $K=6$, 7 and 8 are presented in Figure \ref{fig5:}.
For $K=6$, Figure \ref{fig5:}(a) reveals that different $(\s/l_x,\,m)$ free-parameter choices can result in different vortices not being identified. 
Notably, vortices 3, 5 and 6 are missed in some realizations, which reduces their mean probability and increases uncertainty for those clusters.
Figure \ref{fig5:}(b), for $K=7$, presents six vortices identified with similar probability distributions to the ones obtained by varying $\s$ alone in Section~\ref{subsec:parameter_gaussian} (in Figure \ref{fig3:}(c,d)).
Such similarity between these cases highlights the dominance of the $\s$-sensitivity over the choice of $m$, and the fact that these two parameters do not compound each other. 
Finally, Figure \ref{fig5:}(c) presents the results for $K=8$, highlighting the consistent identification of the jet as a coherent cluster in the system.
The detection of the jet has little impact on the probability distributions for the six vortices.  Both the means and standard deviations for the vortices remain similar to the ones obtained for $K=7$, with a slight standard deviation increase for the cores.
The jet cluster has an overall higher sensitivity to $\sigma$ and $m$, with  higher standard deviations than the vortex cores in Figure~\ref{fig5:}(b).

While these ensemble statistics could be used as a basis for setting the method free-parameters, a thorough investigation on this is beyond the focus of the parameter sensitivity study presented in this paper.  
Based on the results for $K=6$, 7, and 8, it would be reasonable to state that $K=7$ could be chosen over the other options, as it is the case leading to the smallest space-averaged uncertainty.
At the same time, however, the jet is a structure that differentiates itself from the rest of the flow, by behaving in a distinct and coherent way compared to the remaining particles in the chaotic sea.
It could be argued that the choice of $K=8$ for this system is an equally possible way of clustering the system, and provides extra information about the jet cluster, at the price of increasing the result sensitivity to other method free-parameters.
Further work is necessary to automate this selection for a general system.
\section{Ensemble Realizations and Uncertainty to Model Parameters and System Dynamics}
\label{sec:BJ-Ensemble}

In the previous section, a single set of model parameters of the Bickley jet system, with fixed physical parameters, was used to demonstrate how the choice of the method free-parameters impact the clustering results. 
Here, we focus on model parameters that influence the dynamics of the system by changing the velocity field and the resulting trajectories. 
Multiple realizations of the system are used to determine the clustering sensitivity to model parameters, allowing for a characterization of the robustness of the identified clusters.
Section~\ref{subsec:bj_randomization} explains how system parameters are randomized to generate multiple dynamically different realizations.
Section~\ref{subsec:bj_uq} presents the statistics over the described realizations, leading to an uncertainty quantification of the clustering results to model~parameters.

%%%%%%%%%%
%%% Sec 4.1
\subsection{Perturbing the Bickley Jet Dynamics}
\label{subsec:bj_randomization}

While system parameters such as $U$, $L$, $c_2$, and $c_3$ are set by physical arguments (as discussed in Section~\ref{sec:BJ-Central}), the wave amplitudes $A_n$ and phases $\phi_n$ are not, despite exerting a major influence on the system dynamics~\citep{rypina2007lagrangian}.
We use these values as unknown model parameters to introduce variability in the system dynamics.
The amplitudes and phases are drawn from normal distributions centered around the values used in Section~\ref{sec:BJ-Central} (and \citep{hadjighasem16spectral}). 
The realization presented in Section~\ref{sec:BJ-Central} is hereafter referred to as the central realization, and corresponds to amplitudes $A_n=\overline{A}_n$, with $\overline{A}_1=0.0075$, $\overline{A}_2 = 0.15$, and $\overline{A}_3 = 0.30$, and phases $\phi_n=0$. 
Therefore, each realization $I$ in this section is generated by
\eq{
\label{eq:NormalDistribution}
A_n^{(I)}/\,\overline{A}_n \sim \mathcal{N} \left( 1, \, \left(\frac12\right)^2 \right) \quad \textrm{and} \quad
\phi_n^{(I)}/\,l_x \sim \mathcal{N} \left( 0, \, \left(\frac1{24}\right)^2 \right),
}
where $\mathcal{N} \left(\mu,\, \Sigma^2\right)$ denotes a normal distribution of mean $\mu$ and standard deviation $\Sigma$.
The standard deviations for $A_n^{(I)}$ are scaled by the corresponding mean values, while the standard deviation for $\phi_n^{(I)}$ is chosen small enough so that the vortex centers for each realization are likely to be inside of the area covered by the vortices in the central realization.

A total of $R=1000$ realizations are generated from these distributions, and the spectral clustering algorithm described in Section~\ref{sec:Method} is applied with method parameters fixed to $\s/l_x=0.020$, $m=2$ and $K=7$.
While it is possible that individual realizations may require a different method free-parameter selection, our goal in presenting this section is to separate the effects of method free-parameters from those of model parameters.
By sampling model parameters together with method free-parameters, the cluster uncertainty resulting from the model would be obfuscated.
We thus fix the method free-parameters based on the central realization only,  while considering multiple realizations with varying model parameters.

% Fig 06 - a few realizations with different dynamics
\begin{figure}[t]
\centering
\includegraphics[width=\textwidth]{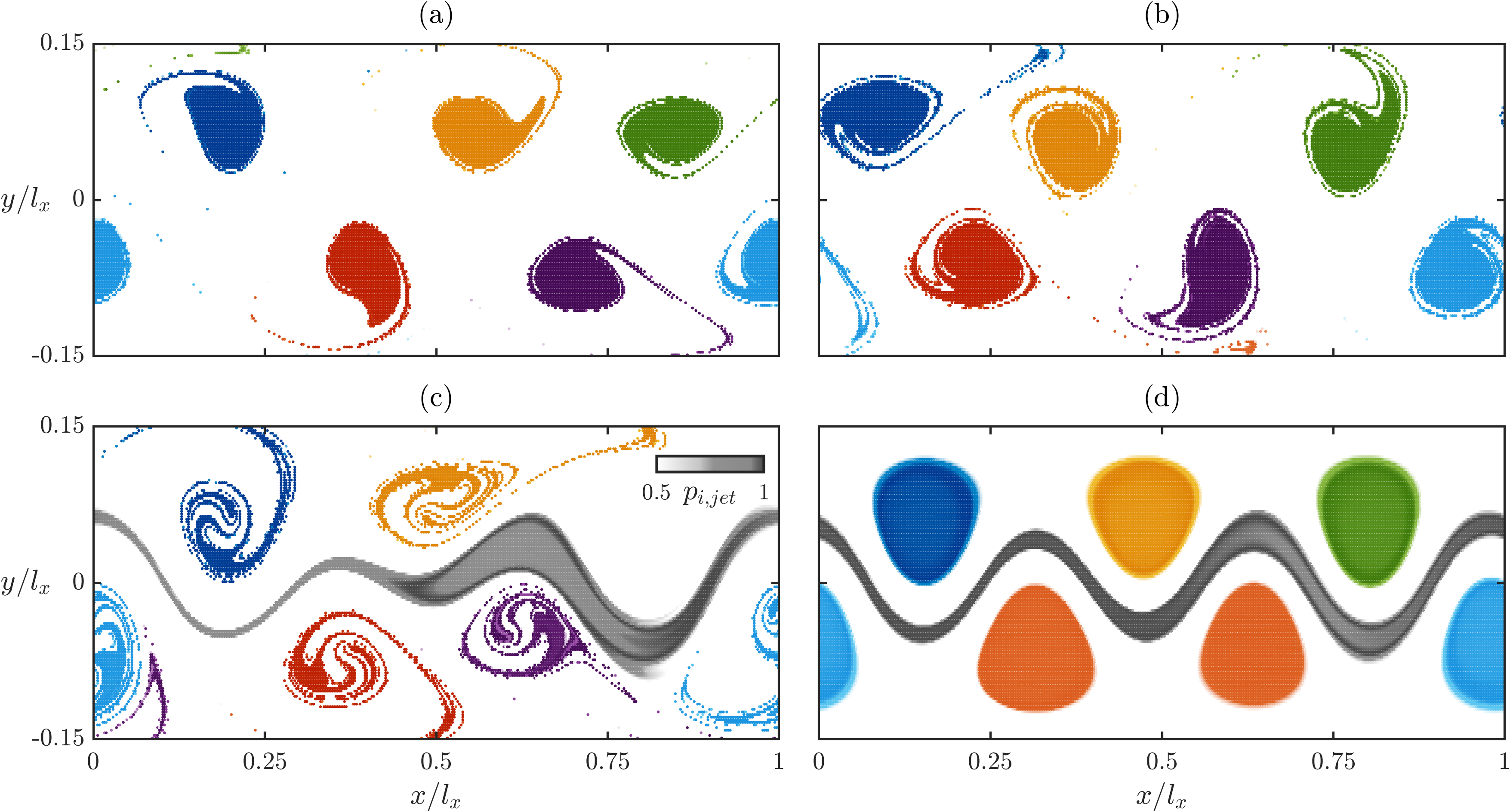}
\caption{Examples of membership probabilities for the six identified clusters, plotted at $t=t_0$, for different model parameters $\{A_n\}$ and $\{\phi_n\}$ drawn from normal distributions.
Cases correspond to parameters ($A_1$, $A_2$, $A_3$, $\phi_1/l_x$, $\phi_2/l_x$, $\phi_3/l_x$) equal to
(\textbf{a}) (0.0087, 0.19, 0.20, 0.01, 0.06, -0.03), (\textbf{b}) (0.0069, 0.26, 0.25, 0.00, -0.08, 0.09), (\textbf{c}) (0.0102, 0.35, 0.25, 0.01, -0.01, 0.01), where the jet is identified and one of the vortices missed, and (\textbf{d}) (0.0077, 0.11, 0.32, 0.00, -0.01, 0.02), where the jet (grayscale) is identified and two vortices are merged. Corresponding colorbars for the vortices presented in Figure~\ref{fig1:hadjighassem}.
(See Supplementary Material for a video of the time evolution of the clustered trajectories.)}
\label{fig6:diff_clusters}
\end{figure}

Figure~\ref{fig6:diff_clusters} presents the membership probabilities for four realizations.
While the realizations do modify the position, shape, and dynamics of the vortices (see videos in Supplementary Material), their presence and number, as well as the presence of the shear jet, mostly remain unchanged.
Figure~\ref{fig6:diff_clusters}(a,b) presents how variable the initial cluster sizes and shapes can be, as well as the effect of the wave phases on the nonuniform spacing between the identified vortices at $t_0$.
While for most realizations all six vortices are identified, there are cases where some of the expected vortices are not identified.
Figure~\ref{fig6:diff_clusters}(c) presents a case in which the jet is identified and one of the vortices is missed.
For that case, vortex 5 gets identified as part of the chaotic sea (not plotted), while a highly asymmetric jet is identified as another coherent cluster in the system, and trajectories are assigned membership probabilities $p_{i,jet}$.
Figure~\ref{fig6:diff_clusters}(d) presents a case for which a more symmetric jet is identified as a cluster and two of the vortices (2 and 4) are merged into a single cluster. 
Other realizations, not illustrated here, result in cases where only a few (or even none) of the vortices are identified. 
For those realizations, there is no clear Eulerian signature of the six vortices.
In what follows, only clusters that identify one and only one vortex are considered for statistical purposes.

%%%%%%%%%%
%%% Sec 4.2
\subsection{Uncertainty Quantification of Ensemble Simulations}
\label{subsec:bj_uq}

% Fig 07 - mean and std for variable dynamics
\begin{figure}
\centering
\includegraphics[width=\textwidth]{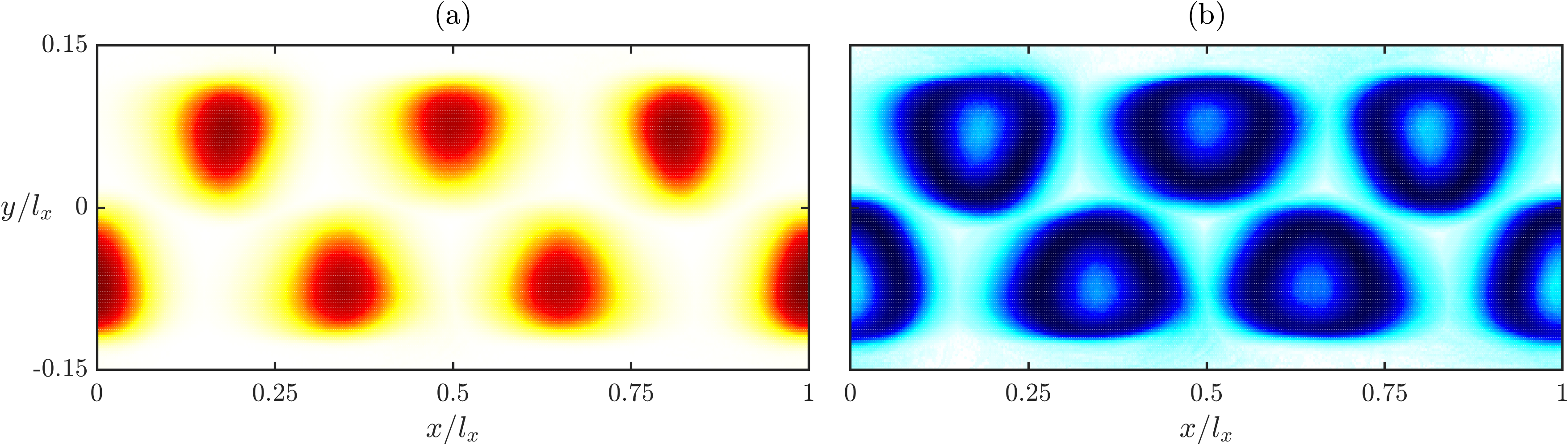}
\caption{Clustering statistics, plotted at $t=t_0$, for an ensemble of $R=1000$ realizations of the Bickley jet system with variable amplitudes $A_n$ and phases $\phi_n$ drawn from normal distributions around the central realization values. Membership probability (\textbf{a}) superimposed mean $\overline{p}_{i,\hat{k}}$ and (\textbf{b}) superimposed standard deviation $S_{i,\hat{k}}$, for vortices $k \in \{1,\ldots,6\}$. Corresponding colorbars presented in Figure~\ref{fig3:}.}
\label{fig7:}
\end{figure}

The clustering results for all $R=1000$ realizations are analyzed to measure their uncertainty statistics.
For each one of the vortices, the mean and standard deviation over the ensemble of parameter value sets of the Bickley jet are computed according to \eqref{eq:mean} and \eqref{eq:std}.
Figure~\ref{fig7:} presents the superimposed membership probability mean, $p_{i,\hat{k}}$, and corresponding superimposed sample standard deviation, $S_{i,\hat{k}}$, for the ensemble.
Even with a variety of behaviors observed in the ensemble, as illustrated in Figure~\ref{fig6:diff_clusters}, averaging the ensemble smooths the vortices, resulting in cluster shapes and positions that are similar to the ones obtained for the central realization (Figure~\ref{fig1:hadjighassem}(a)).
However, the mean probabilities in Figure~\ref{fig7:}(a) highlight how introducing uncertainty in the wave amplitudes and phases leads to smaller vortex cores with high membership probabilities.
The mean probabilities now peak at 0.91 rather than 1.00, due to the phase shift between realizations.
The membership probability decay from the vortex cores to the perimeters is also more gradual than for the central realization, and the averaging clears out previously identified filaments that are realization-dependent.

Figure~\ref{fig7:}(b) highlights a more spread out distribution and overall higher magnitude for the superimposed standard deviation.
Moreover, higher standard deviations are now observed at the vortex cores, which arises from the phase parameter uncertainty.
Also, notice that most of the low uncertainty regions associated to the jet in Figure~\ref{fig3:}(b) have higher uncertainty in Figure~\ref{fig7:}(b).
With the varying model parameters and dynamics, the vortex positions, shapes and trajectories are more variable.
All of these introduce uncertainties that are not observed for the central set of parameter values.
The ensemble analysis, therefore, highlights structure sensitivities (and robustness) that are not apparent from the central realization alone.

% Fig 08 - uncertainty demonstration
\begin{figure}[!t]
\centering
\includegraphics[width=1\textwidth]{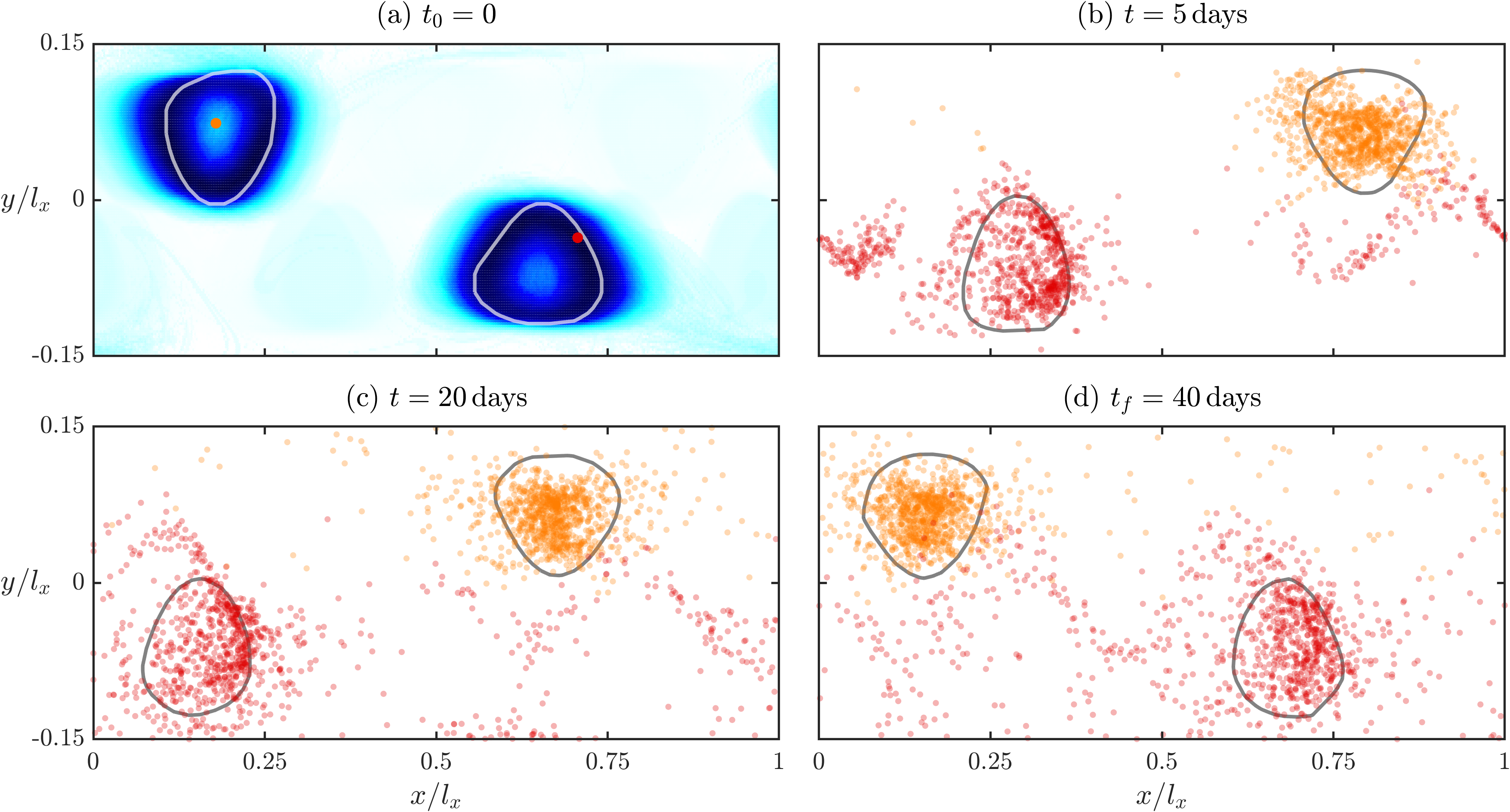}
\caption{Evolution of trajectories from the 1000 realizations released from a low (orange, top) and high (red, bottom) uncertainty position. (\textbf{a}) Orange particles are initialized at the core of vortex 1, while red particles are initialized at the perimeters of vortex 4. Positions are plotted on top of the superimposed standard deviation $S_{i,\hat{k}}$ computed using vortices 1 and 4, and plotted at time $t_0=0$.
The positions for the multiple realizations and the central realization vortex boundaries are presented at (\textbf{b}) $t=5$, (\textbf{c}) $t=20$ and (\textbf{d}) $t_f=40$ days.
Transparency is used to highlight high or low concentration of particles. The vortex boundaries from the central realization are plotted in gray and enclose particles with membership probabilities greater than 0.5, for clusters 1 and 4. Corresponding colorbar for (\textbf{a}) presented in Figure~\ref{fig3:}.
(See Supplementary Material for a video of the time evolution.)}
\label{fig8:demonstration}
\end{figure}

To illustrate the ramifications of regions of high uncertainty, we demonstrate how different ensemble trajectories are when initialized at high and low uncertainty locations.
At $t_0=0$, the orange particles in Figure~\ref{fig8:demonstration}(a) are initially located at the core of cluster 1 and correspond to $\overline p_{i,1}=0.883$ and $S_{i,1}=0.287$, while the red particles start at the perimeter of cluster 4 and correspond to $\overline p_{j,4}=0.469$ and $S_{j,4}=0.475$.
Figure~\ref{fig8:demonstration}(b,c,d) presents snapshots of the particle positions for all 1000 realizations at three different times.
For reference, the gray boundaries enclose vortices $k=1$ and $4$ for the central realization (for particles such that $p_{i,k}\geq 0.5$).
While particles released in the position of lowest uncertainty (orange) remain concentrated, particles released in the position of highest uncertainty (red) quickly spread throughout the domain.
Particle concentration is more pronounced for the case of lower uncertainty, with 76.4\% of the particles remaining inside the corresponding boundary after $t=40$ days, as opposed to only 45.0\% for the higher uncertainty case. 
The presence of the jet separating the top and bottom of the domain keeps most orange and red particles from moving to the opposite half.

Regions of higher uncertainty, not characterized when considering the central realization only, are revealed by the ensemble analysis, and correspond to flow regions for which particle cluster membership is most unknown.  
Our analysis shows that the cores of the vortices are robust even if the model parameters are varied, but the narrow filaments identified in the central realizations should be viewed as less robust as demonstrated by the ensemble analysis.
Further, while both particles in Figure~\ref{fig8:demonstration} are initialized and remain inside of their respective clusters for the central realization, the same happens for less than half of the realizations considered, once model parameter uncertainty is incorporated and accounted for.

\section{Martha's Vineyard Ensemble Forecast and Surface Drifter Trajectories}
\label{sec:MV}

Having demonstrated the clustering uncertainty quantification method on an analytic system, we apply the technique to a real ensemble forecast of a nested primitive equation ocean model, with intrinsic model uncertainty.
The model is used to forecast the three-dimensional velocity field for the coastal region near Martha’s Vineyard, an island located south of Cape Cod, Massachusetts, USA.
Trajectories from drifters deployed~\citep{filippi2020parameter-free} for the corresponding day are then compared to the cluster behaviors.
Section~\ref{subsec:mseas_drifters} presents characteristics of the Martha's Vineyard region, the model used to forecast the velocity field, and the results of the clustering uncertainty quantification analysis applied to the ensemble forecast.
Section~\ref{subsec:uq_MV} details the drifter experiment and compares the drifter trajectories to the forecast trajectory clustering results and the associated uncertainty.

\subsection{Velocity Model Ensemble Forecast and Uncertainty Quantification}
\label{subsec:mseas_drifters}

The island of Martha's Vineyard, with an area of almost \SI{250}{km^2}, is the largest island in New England, and lies \SI{11}{km} off the coast of Cape Cod.
The prevailing currents in the coastal region south of the island are associated with wind-driven coastal divergence, tidal forcing and a mean southward drift~\citep{rypina2014eulerian}.
During the summer months, the region experiences a mean westward surface current that reaches velocities of \SI{15}{cm\, s^{-1}}~\citep{rypina2016investigating}.
The drifter deployment experiments targeting predicted coherent structures~\citep{filippi2020parameter-free}, presented in Section~\ref{subsec:uq_MV}, took place around the \SI{2.5}{km^2} uninhabited island of Nomans Land, south of Martha's Vineyard.
The channel between the two islands has a width of approximately $\SI{5}{km}$ and an average depth of \SI{10}{m}.

We used the MIT Multidisciplinary Simulation, Estimation, and Assimilation Systems (MIT-MSEAS) primitive equation ocean modeling system~\citep{haley2010multiscale, haley2015optimizing} to compute ocean surface velocity forecasts in the Martha’s Vineyard coastal region during August 2018.
The modeling system provided forecasts of the ocean state variable fields (three-dimensional velocity, temperature, salinity, and sea-surface height) every hour, with a spatial resolution of \SI{200}{m}.
More details about the model forecast initialization, tidal forcing, atmospheric flux forcing, and CTD data assimilation can be found in~\citep{serra2020search, filippi2020parameter-free}.
The deterministic two-way nesting ocean forecast initialized from the estimated ocean state conditions at a particular time is referred to as the central forecast, and ensemble forecasts were initialized using Error Subspace Statistical Estimation procedures~\citep{lermusiaux2002mapping}.
The forecasts within the ensemble were initialized from perturbed initial conditions of all state variables and forced by perturbed tidal forcing, atmospheric forcing fluxes and lateral boundary conditions.
These perturbations were created to represent the expected uncertainties in each of these quantities.
Finally, parameter uncertainties (bottom drag, mixing coefficients, etc.) were also modeled by perturbing the values of parameters for each forecast.

\begin{figure}[t]
\centering
\includegraphics[width=\textwidth]{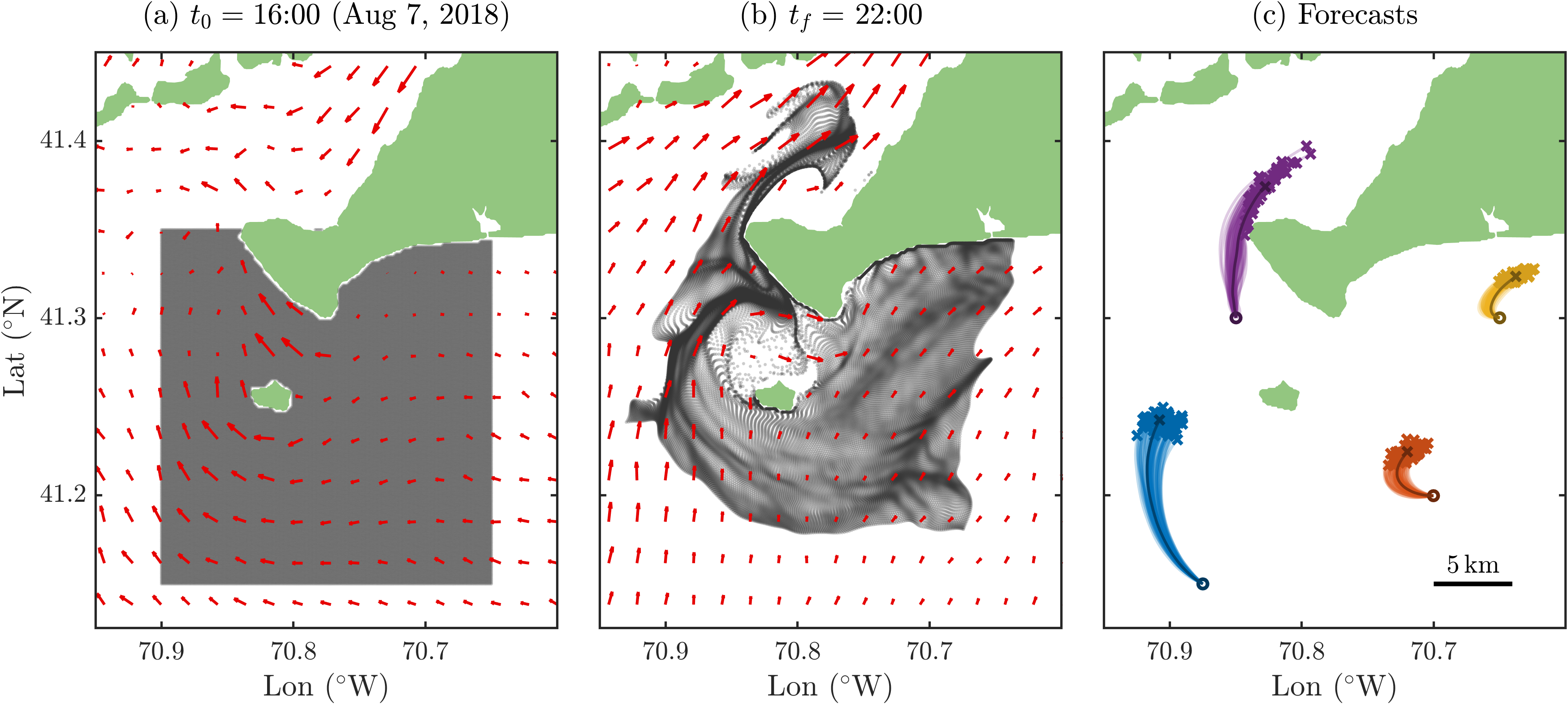}
\caption{Martha's Vineyard coastal area, central forecast particle distribution, and superimposed forecast model velocity field at (\textbf{a}) the initial time $t_0=16$:00, (\textbf{b}) the final time $t_f=22$:00~UTC. (\textbf{c})~Trajectories obtained in each of the 71 forecasts for 4 different initial positions. Circles represent the initial positions and crosses the final positions, after 6 hours. The darker trajectories correspond to the central forecast.}
\label{fig9:}
\end{figure}

A total of 71 forecasts were considered for the present study.
Fluid particle trajectories confined to the surface were analyzed over a \SI{6}{h}-time-window, between $t_0=16$:00 and $t_f=22$:00~UTC on August~7, 2018.
Such a short timing is critical in search-and-rescue operations, as after six hours the likelihood of rescuing people alive drops significantly~\citep{serra2020search}.
The forecast velocity fields used for trajectory integration were generated the night before the experiment.
%\footnote{data available at \url{http://transport.me.berkeley.edu/thredds/catalog/public/MIT-MSEAS/2018-08-FullStudy/2018-08-06/Ocean_2D_Currents_At_Surface_Multiple_Ensembles/catalog.html})}.
Synthetic trajectories were computed using the web-based gateway Trajectory Reconstruction and Analysis for Coherent Structure Evaluation~\citep{ameli2019transport, trace}.
At time $t_0$, particles are uniformly distributed on a 250-by-250 grid covering the domain $[70.65^\circ$W$,70.90^\circ$W]$\times [41.15^\circ$N$,41.35^\circ$N], from which portions corresponding to land are removed.
This grid is approximately $\SI{21}{km}$ by $\SI{22}{km}$.

Figure \ref{fig9:}(a) presents the initial particle distribution, superposed with the velocity field for the central forecast.
Trajectories are integrated using an adaptive 7\textsuperscript{th}-order Runge-Kutta-Fehlberg method and bicubic spline spatial interpolation, with free-slip boundary conditions applied near land.
Particle positions are output every 5 minutes.
Figure \ref{fig9:}(b) presents the final positions of the particles for the central forecast.
Darker regions correspond to particles collecting, which is mostly observed along the coast.
The model velocity field captures the reversal of the tide, as can be observed from the flipping in the average flow direction between $t_0$ in Figure \ref{fig9:}(a) and $t_f$ in Figure \ref{fig9:}(b).
Trajectories obtained in each of the 71 forecasts are presented in Figure \ref{fig9:}(c) in contrast with the central forecast, for particles initialized at distinct positions, to demonstrate the degree of trajectory variability between forecasts.

\begin{figure}[t]
\centering
\includegraphics[width=1\textwidth]{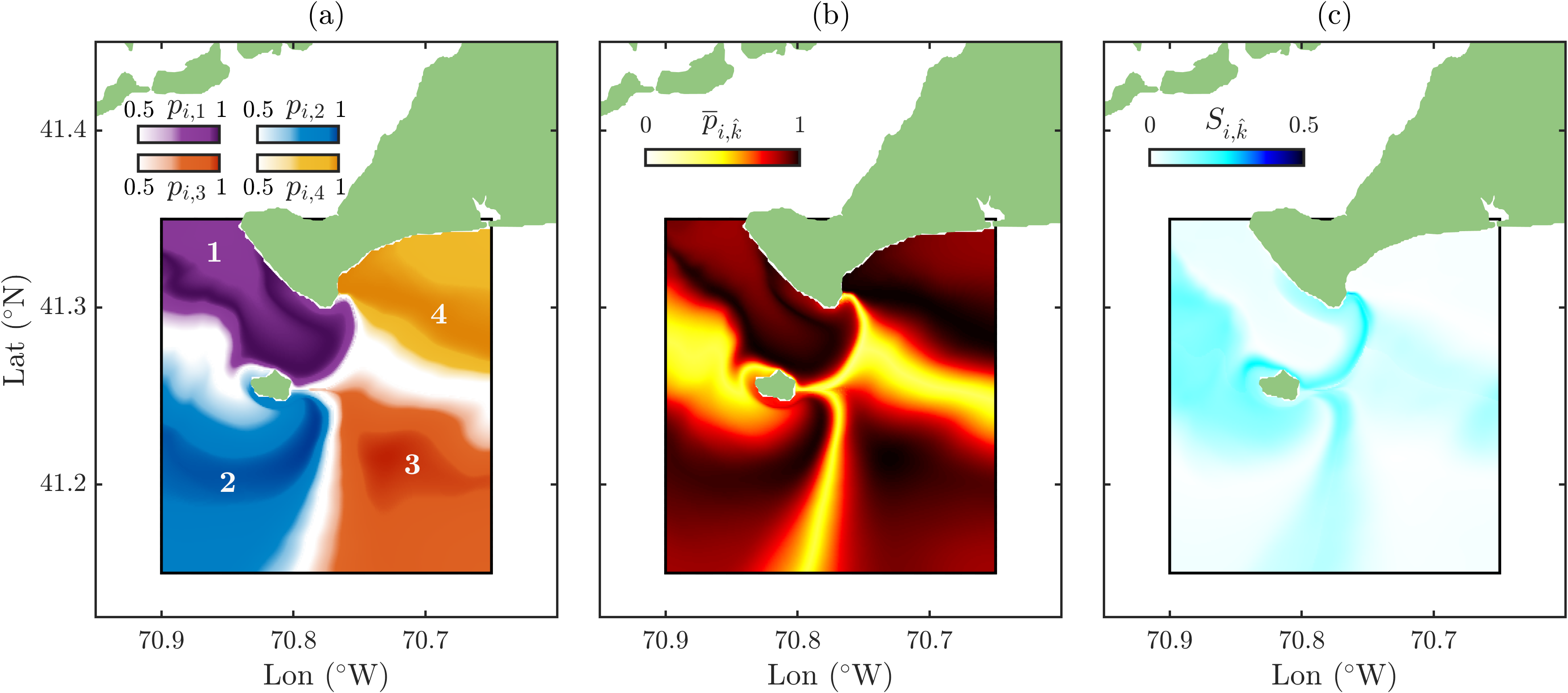}
\caption{(\textbf{a}) Central forecast membership probabilities at $t=t_0$, for clusters ${k \in \{1,\ldots,4\}}$. The black box encloses the domain where trajectories are initialized. Clustering statistics for ensemble of $R=71~$forecasts: superimposed membership probability (\textbf{b}) mean $\overline{p}_{i,\hat{k}}$ and (\textbf{c}) standard deviation $S_{i,\hat{k}}$. The color maps presented in (\textbf{a}) are also used in Figure \ref{fig11:}.}
\label{fig10:}
\end{figure}

A similar study to the one in Section~\ref{sec:BJ-Central} was performed for the central forecast, to determine the spectral clustering sensitivity to method free-parameters.
We select the following method free-parameters: similarity radius $\s=\SI{1}{km}$, fuzziness $m=2$ and number of clusters $K=4$, which are used for all $R=71$ forecasts.
While it is possible to break the domain into fewer clusters, four clusters minimize the space-averaged standard deviation of the results.
Because the graph is fully connected with this choice of $\s$, $\lambda_1=0$ and the components of $\mathbf{q}_1$ are constant~\citep{vonluxburg07}.
Therefore, the clustering (in the absence of a chaotic sea) into $K=4$ clusters is performed using the information from eigenvectors $\mathbf{q}_2,\mathbf{q}_3$ and $\mathbf{q}_{4}$ only.
The membership probabilities for the central forecast are presented in Figure~\ref{fig10:}(a) for trajectories with $p_{i,k}\geq0.5$.
The domain is partitioned into 4 quadrants of similar size, and gaps between clusters correspond to particles with membership probabilities lower than $0.5$.

Different forecasts are used to compute the means $\overline{p}_{i,k}$ and sample standard deviations $S_{i,k}$ for ${k \in \{1,\ldots,4\}}$.
The superimposed mean $\overline{p}_{i,\hat{k}}$ and standard deviation $S_{i,\hat{k}}$ are presented in Figure~\ref{fig10:}(b,c).
The parameterization used for the model produces only a modest variation in the trajectory outcomes over six hours as demonstrated in Figure~\ref{fig9:}(c).
Regions of highest uncertainty for this system correspond to identifying the edges of the clusters accurately, but this level of uncertainty is significantly lower than those observed in the Bickley jet example.
The most pronounced uncertainty regions, in Figure~\ref{fig10:}(c), correspond to the boundary between clusters 1 and 4, followed by that between clusters 1 and 2.

\subsection{Drifter Data and Forecast Cluster Dynamics}
\label{subsec:uq_MV}

\begin{figure}[!t]
\centering
\includegraphics[width=0.667\textwidth]{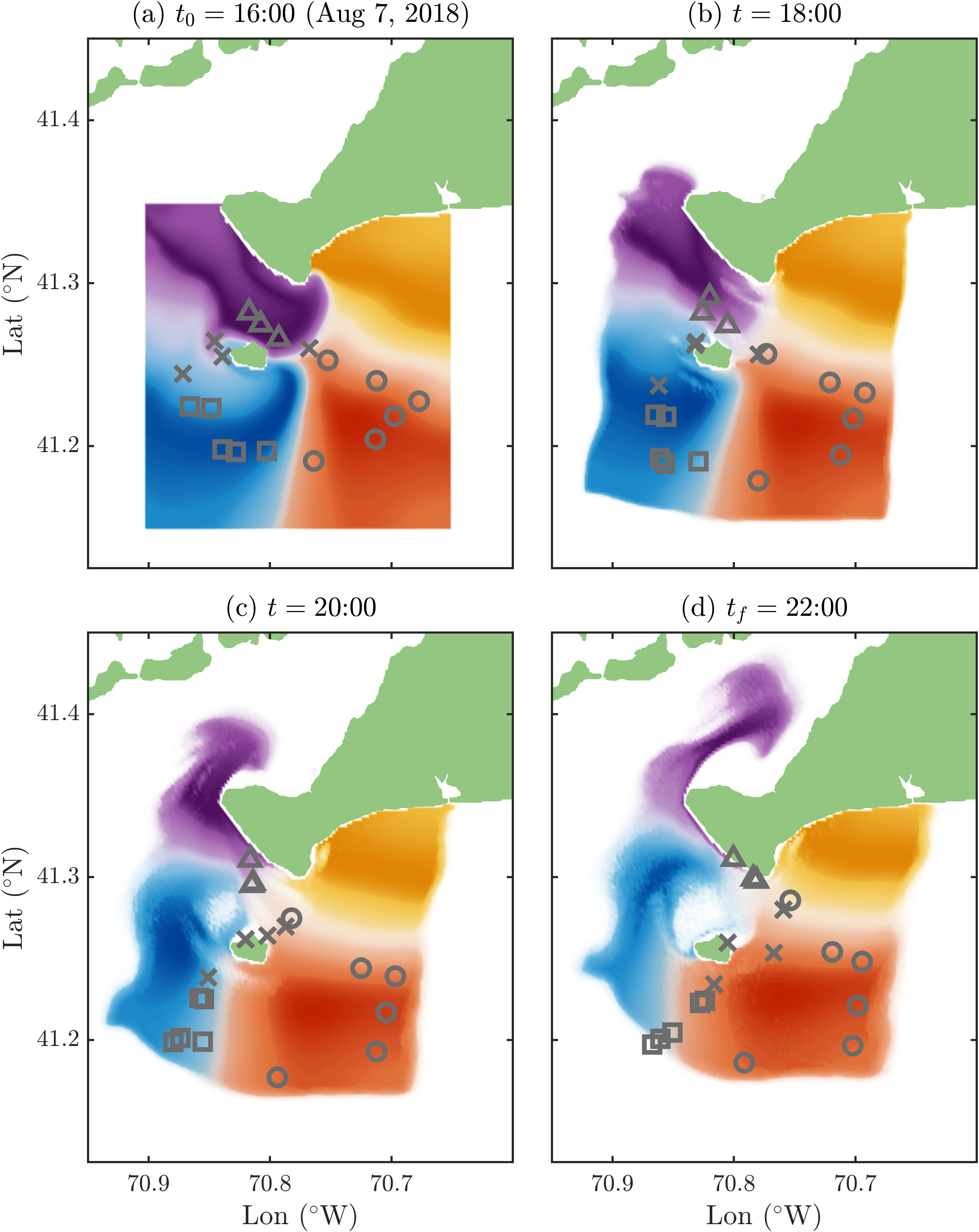}
\caption{Time evolution of the average clusters in a binned domain. The 18 drifters are marker-coded depending on the cluster in which they started. Crosses represent the 4 drifters initialized at locations of higher uncertainty, with $S_{i,\hat{k}}>0.1$. The times correspond to (\textbf{a}) $t_0=16$:00, (\textbf{b}) $t=18$:00, (\textbf{c}) $t=20$:00, and (\textbf{d}) $t_f=22$:00~UTC.
Triangles are initialized on the purple cluster (3 drifters), squares on the blue cluster (5 drifters) and circles on the orange cluster (6 drifters). Corresponding colorbars presented in Figure~\ref{fig10:}(a). 
(See Supplementary Material for a video of the time evolution.)}
\label{fig11:}
\end{figure}

The experiment targeting predicted coherent structures consisted  of surface drifter releases that took place on August 7, 2018, in the vicinity of Nomans Land ~\citep{filippi2020parameter-free}.
The CODE drifters used in the experiment have technical specifications listed in~\citep{serra2020search,filippi2020parameter-free}.
Drifters of the same design are routinely used by the U.S. Coast Guard in search-and-rescue operations, as well as in previous field experiments in the coastal region near Martha’s Vineyard~\citep{rypina2014eulerian,rypina2016investigating}.
The drifters were equipped with GPS transmitters that provided positioning fixes every \SI{5}{min}, with an accuracy of a few meters.
An elliptical route around Nomans Land was used for the drifter deployment, employing two WHOI vessels to minimize ship time, so that all drifters were in water by the start of the interval of analysis.
Eighteen drifters were deployed in the water around the predicted locations of the targeted coherent structures.
The position of the drifters at the starting time of our analysis, $t_0=16$:00\, UTC, is presented in Figure~\ref{fig11:}(a).

Figure \ref{fig11:}(a-d) presents the mean cluster trajectories for each of the 4 clusters, on which we superpose the drifter positions from~\citep{filippi2020parameter-free}.
To plot the evolution of $\overline{p}_{i,k}$, the entire domain is first split into $175\times175$ bins. 
Then, at each time instance, an average membership probability for each bin, over the forecasts, is calculated.
This average is based on particles inside each bin at time $t$, and weighted by their membership probabilities $p_{i,k}$.
The overall system dynamics are similar to what was observed for the central forecast in Figure~\ref{fig9:}(a,b), with the clusters highlighting groups of different behavior.
Using the labels from Figure~\ref{fig10:}(a), while cluster 1 (purple) travels a longer distance to the northeast, ending north of Martha's Vineyard, cluster 2 (blue) moves a shorter distance northward, while clusters 3 and 4 (red and yellow) are less dynamic and remain mostly to the east of Nomans Land.

Drifters are marker-coded according to the mean membership probabilities $\overline{p}_{i,k}$ of their spatial position at $t_0$.
Triangles correspond to cluster 1, squares to cluster 2, circles to cluster 3, and no drifters were initialized in cluster 4.
Drifters initialized in higher uncertainty regions (see Figure~\ref{fig10:}(c)) are plotted as crosses, and correspond to regions where $S_{i,\hat{k}}>0.1$, or $\overline{p}_{i,\hat{k}}<0.7$.
The results demonstrate that the drifters predominately remain inside of the forecast model clusters during the first four hours of the time interval, and exceptions to this behavior are associated with higher uncertainty.
Consider, for example, the three drifters represented as crosses, initially west of Nomans Land.
Two of them move along the northern coast of the island, with one of them beaching and the other ending on cluster 2.
The third drifter headed south first, then east, also toward cluster 2.
This fact highlights the value of the uncertainty quantification analysis to understand different dynamical behaviors in the flow which are not captured by the central forecast alone.
All drifters were eventually advected eastward after~20:00.

Some discrepancies between the clustering predicted behavior and the drifter trajectories were observed.
Wind gusts that occurred between 16:00 and 20:00 significantly affected the drifter trajectories.
These gusts had not been predicted by the forecasts from the National Centers for Environmental Prediction used by the MIT-MSEAS model for the atmospheric forcing~\citep{filippi2020parameter-free}.
The study was therefore limited by differences between observed flows and model flows, and the coherent structure analysis accuracy is limited to the accuracy of the velocity field used for processing.
Nonetheless, drifters released in the clusters with high membership probabilities
% ended up describing similar trajectories, regardless of model inaccuracies, even if they did not precisely match the predicted cluster behavior.
tend to behave similarly to the clusters, even if they do not precisely match the predicted trajectory behavior.
This fact highlights the coherent structure robustness, even in uncertain conditions.
The clustering analysis partitions the domain into robust regions where a drifter released in the region will remain there or nearby over the period of analysis.
The uncertainty quantification analysis helped identifying the key structures delimiting regions with different transport behaviors, further showing and expanding the applicability of trajectory clustering for studying oceanic flows, despite model imperfections.

\section{Conclusions}
\label{sec:Conclusions}

Ensemble statistics of the trajectory clustering results provide a partitioning of the fluid domain that may provide critical information in emergency response situations, such as search-and-rescue operations, when operational decisions about optimal resource allocation need to be made quickly, accurately, and account for model uncertainties.
We presented a modified version of the spectral clustering method with soft membership probabilities, and applied it to fluid particle trajectories to identify coherent structures first in an analytic flow model, then in forecast simulations provided by a coastal ocean model.
Uncertainty quantification was applied to assess both the result sensitivity to the clustering method free-parameters and the cluster variability with unknown parameters of the model data.
The method sensitivity study, performed on the analytic quasi-periodic Bickley jet system, identified the similarity radius as the free-parameter to which the clusters are most sensitive.
To mimic model uncertainty, the Bickley jet parameters were varied to perform a model sensitivity study that highlights the robustness of vortex cores compared to the more uncertain vortex perimeters. 

Finally, the method was applied to an ocean ensemble forecast of the coastal region of Martha's Vineyard, and the clustering results were compared to drifter trajectories from a drifter release experiment targeting predicted coherent structures.
% Forecast cluster behaviors were confirmed by drifter trajectories, as drifters deployed within each cluster performed similar motions to those predicted from the forecast velocity model.
The forecast clusters from the ensemble analysis provided a good baseline for the drifter behavior, as drifters deployed within each cluster performed similar motions to their corresponding clusters.
Ocean transport predictions are challenging due to the complexity of the underlying dynamics governing the flow, and while the Lagrangian approach ultimately depends on the accuracy of the available velocity fields and the quality of the model data, the results presented here demonstrate that the identified clusters are robust to uncertainties in the model and able to predict the main elements of the organization of flow transport. 
The coupling of the clustering approach to uncertainty quantification can provide a more complete and informative description of flow transport and areas of higher and lower uncertainty within different clusters.
Despite not having been the case for the drifter release presented here, a clustering forecast analysis could be used in planning a drifter deployment for future experiments.

Further refinement of the trajectory clustering method is highly desirable, in particular aiming to reduce parameter sensitivity.
The uncertainty quantification with respect to method free-parameters could be used to select parameters that minimize cluster variability, in order to identify clusters that are physical structures, and not a byproduct of the system and parameters chosen.
On a more fundamental level, while the method presented here provides robust clusters, it may be possible to improve the method by incorporating fundamental changes to the similarity measure, rather than addressing the sensitivity to free-parameters only.
To quantify trajectory similarity, one could not only consider the similarity between the particle spatial coordinates in time, but also that of their velocity vectors, and propose a hybrid notion of similarity.
Finally, applying the method to other oceanic forecasts and using the forecast clustering results to plan and execute drifter release experiments can be a promising path to more effective experiments, by increasing the likelihood of released drifters capturing targeted coherent structures and ocean transport barriers.

\section*{Author Contributions}
Conceptualization, G.S.V. and M.R.A.; Formal analysis, G.S.V.; Funding acquisition, I.I.R.; Methodology, G.S.V. and M.R.A.; Software, G.S.V.; Supervision, M.R.A.; Visualization, G.S.V.; Writing – original draft, G.S.V.; Writing – review \& editing, G.S.V., I.I.R. and M.R.A.

\section*{Funding}
This research was funded by National Science Foundation Division of Atmospheric and Geospace Sciences, award number 1520825.

%%%%%%%%%%%%%%%%%%%%%%%%%%%%%%%%%%%%%%%%%%%%%%%%%%%%%%%%%%%%%%%%%%%%%%%%%%%%%%%%
% ACKNOWLEDGMENT
%%%%%%%%%%%%%%%%%%%%%%%%%%%%%%%%%%%%%%%%%%%%%%%%%%%%%%%%%%%%%%%%%%%%%%%%%%%%%%%%
\section*{Acknowledgments}
We thank Margaux Filippi and Thomas Peacock for sharing the drifter data set, Pierre Lermusiaux for providing the forecast ensemble, and H. M. Aravind for his helpful suggestions.

\section*{Conflicts of Interest}
The authors declare no conflict of interest.

%%%%%%%%%%%%%%%%%%%%%%%%%%%%%%%%%%%%%%%%%%%%%%%%%%%%%%%%%%%%%%%%%%%%%%%%%%%%%%%%
%%% REFERENCES
%%%%%%%%%%%%%%%%%%%%%%%%%%%%%%%%%%%%%%%%%%%%%%%%%%%%%%%%%%%%%%%%%%%%%%%%%%%%%%%%
% \clearpage
% \nocite{*} % Comment this line to show only the ones cited in the text
\bibliography{references.bib}

\appendix

% %%%%%
% %%% APPENDIX A
\clearpage
% \clearpage
\pagestyle{empty}
\section{Similarity Matrix Sparsification}
\label{appendix:A}

As mentioned in Section \ref{subsec:bickley_and_spectral}, the use of the Gaussian similarity measure
\eq{
\label{eqA:Gaussian}
w_{ij} = \exp\left(- r^2_{ij}/2\sigma^2\right)
}
in the similarity matrix $\mathbf{W}$ greatly reduces the impact of the sparsification step compared to the $l_x/r_{ij}$ measure used in~\citet{hadjighasem16spectral}.
This result is due to $w_{ij}$ in \eqref{eqA:Gaussian} approaching zero faster for large $r_{ij}$.

It is mentioned in Section \ref{subsec:parameter_gaussian}, however, that sparsifying the matrix is worthwhile to reduce computational costs and storage, and we therefore sparsify entries of $\mathbf{W}$ that satisfy ${w_{ij}<\exp(-4^2/2)}$, corresponding to $r_{ij} > 4\sigma$.
This sparsification rule is used in Sections \ref{subsec:parameter_sigma}, \ref{subsec:parameter_m}, \ref{subsec:parameter_K} and \ref{subsec:bj_randomization}, under the assumption that such sparsification has negligible impact on the clustering membership probability results.

Here, we present a study for the Bickley jet system with model parameters as in Section~\ref{sec:BJ-Central} and method free-parameters $\s/l_x=0.020$, $m=2$, and $K=7$, where the clustering is performed using different levels of sparsification.
We vary the sparsification levels by setting $w_{ij}=0$ for $r_{ij}/\sigma > \alpha$, with $\alpha\in\{0.5,1,2,4,8\}$ and $\alpha=\infty$ representing the non-sparsified result matching Figure~\ref{fig1:hadjighassem}(a).

\begin{figure}[H]
\centering
\includegraphics[width=\textwidth]{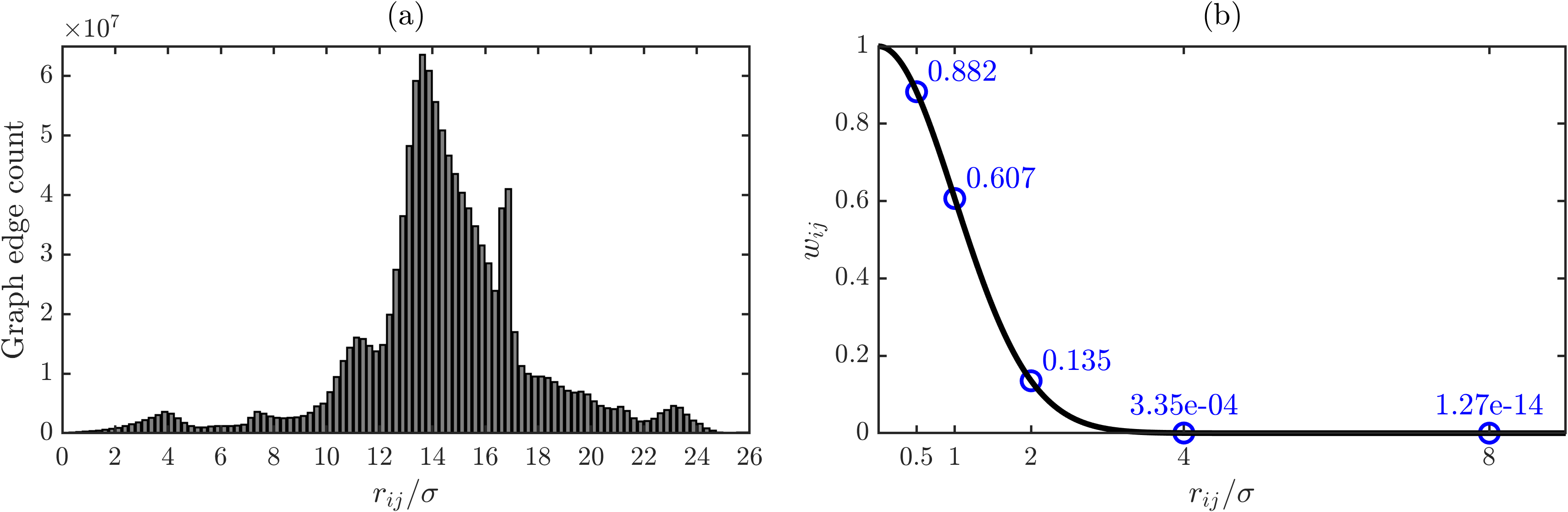}
\caption{(\textbf{a}) Histogram of the time-averaged distance values, made dimensionless by $\s=0.020/l_x$, with a bin width of 0.25. (\textbf{b}) Weight dependence on $r_{ij}/\s$ for the Gaussian measure \eqref{eqA:Gaussian}. The $w_{ij}$ values for $r_{ij}/\s=0.5$, 1, 2, 4 and 8 are shown in blue.}
\label{figA1:}
\end{figure}

Figure~\ref{figA1:}(a) presents a histogram with the distribution of $r_{ij}$ values.
The vast majority of these values represent the time-averaged distance between particles that describe dissimilar trajectories, and therefore correspond to graph edges of negligible weights.
Figure~\ref{figA1:}(b) presents the weight dependence on $r_{ij}/\s$, and therefore the maximum magnitude of matrix entries that are dropped when $\mathbf{W}$ is sparsified for a choice of $\alpha$.

The impact of the sparsification parameter $\alpha$ on the number of retained (nonzero) entries in the similarity matrix is shown in Table~\ref{tabA1:}.
For all finite choices of $\alpha$ presented here, the percent sparsification of $\mathbf{W}$ is greater than 95\%.

\begin{table}[H]
\caption{Dependence of $\mathbf{W}$ entries on the sparsification parameter $\alpha$.}
\label{tabA1:}
\centering
%% \tablesize{} %% You can specify the fontsize here, e.g., \tablesize{\footnotesize}. If commented out \small will be used.
\begin{tabular}{ccc}
\toprule
\textbf{Sparsification parameter $\alpha$}	& \textbf{Number of nonzero entries in $\mathbf{W}$}	& \textbf{Percent Sparsification (\%)}\\
\midrule
0.5 & 287,852 & 99.99\\
1 & 1,078,658 & 99.95\\
2 & 5,052,598 & 99.78\\
4 & 38,712,302 & 98.32\\
8 & 99,413,096 & 95.69\\
$\infty$ & 2,304,000,000 & 0\\
\bottomrule
\end{tabular}
\end{table}

\begin{figure}[H]
\centering
\includegraphics[width=\textwidth]{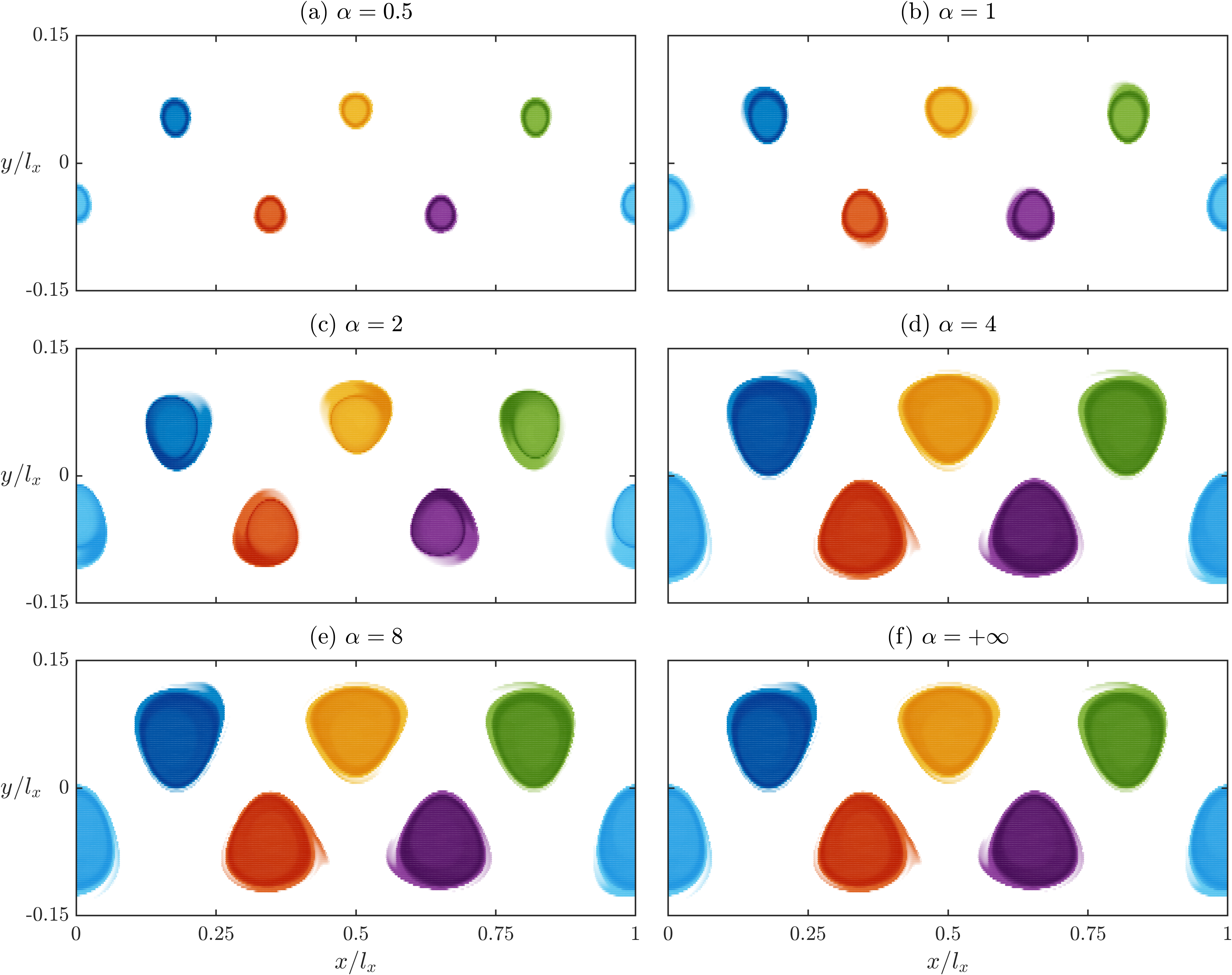}
\caption{Membership probabilities for vortices $k \in \{1,\ldots,6\}$, for different choices of the sparsification parameter $\alpha$, plotted at $t=t_0$. Values correspond to
(\textbf{a}) $\alpha=0.5$, (\textbf{b}) $1$, (\textbf{c}) $2$, (\textbf{d}) $4$, (\textbf{e}) $8$, and (\textbf{f}) $+\infty$. Corresponding colorbars presented in Figure~\ref{fig1:hadjighassem}}
\label{figA2:}
\end{figure}

The cluster membership probabilities obtained with the different choices of $\alpha$ are presented in Figure~\ref{figA2:}.
We notice that sparsifying with $\alpha=0.5$, 1 or 2 clearly has an effect on the resulting membership probabilities, as clusters shrink as $\alpha$ is reduced.
For $\alpha\geq4$, however, the probabilities no longer depend on $\alpha$, and no difference is observed between the method output for $\alpha=4$ or the non-sparsified case ($\alpha=\infty$).

Sparsifying with $\alpha=4$  also reduces the number of nonzero entries in $\mathbf{W}$ by about 60 times, and therefore facilitates storage and reduces the computational time when solving the generalized eigenproblem.
Note that this result is valid for the Bickley jet parameterized as in Section 3, but the impact of sparsification may vary for an arbitrary flow.

% % %%%%%
% % % APPENDIX B - Variable pycnocline cEin
% % %%%%
\clearpage
\clearpage
\section{Video Description}
\label{appendix:Videos}

The four supplementary videos, for which captions are available below, correspond to the time evolution of the results presented in Figures \ref{fig2:sigma_representatives}, \ref{fig6:diff_clusters}, \ref{fig8:demonstration} and \ref{fig11:}.

\begin{itemize}
    \item[$\star$] \texttt{video02.avi}\\
    Membership probabilities for vortices $k \in \{1,\ldots,6\}$, for different choices of the parameter $\s$. Values correspond to
    (\textbf{a}) $\sigma/l_x=0.005$, (\textbf{b}) $0.015$, (\textbf{c}) $0.030$ and (\textbf{d}) $0.040$. Corresponding colorbars presented in Figure~\ref{fig1:hadjighassem}.
    \\
    \item[$\star$] \texttt{video04.avi}\\
    Examples of membership probabilities for the six identified clusters, for different model parameters $\{A_n\}$ and $\{\phi_n\}$ drawn from normal distributions.
    Cases correspond to parameters ($A_1$, $A_2$, $A_3$, $\phi_1/l_x$, $\phi_2/l_x$, $\phi_3/l_x$) equal to
    (\textbf{a}) (0.0087, 0.19, 0.20, 0.01, 0.06, -0.03), (\textbf{b}) (0.0069, 0.26, 0.25, 0.00, -0.08, 0.09), (\textbf{c}) (0.0102, 0.35, 0.25, 0.01, -0.01, 0.01), where the jet is identified and one of the vortices missed, and (\textbf{d}) (0.0077, 0.11, 0.32, 0.00, -0.01, 0.02), where the jet (grayscale) is identified and two vortices are merged. Corresponding colorbars presented in Figures~\ref{fig1:hadjighassem} and \ref{fig4:}.
    \\
    \item[$\star$] \texttt{video08.avi}\\
    Evolution of trajectories from the 1000 realizations released from a low (orange, top) and high (red, bottom) uncertainty position. Orange particles are initialized at the core of vortex 1, while red particles are initialized at the perimeters of vortex 4.
    Transparency is used to highlight high or low concentration of particles. The vortex boundaries from the central realization are plotted in gray and enclose particles with membership probabilities greater than 0.5, for clusters 1 and 4.
    \\
    \item[$\star$] \texttt{video11.avi}\\
    Time evolution of the average clusters in a binned domain. The 18 drifters are marker-coded depending on the cluster in which they started. Crosses represent the 4 drifters initialized at locations of higher uncertainty, with $S_{i,\hat{k}}>0.1$.
    Triangles are initialized on the purple cluster (3 drifters), squares on the blue cluster (5 drifters) and circles on the orange cluster (6 drifters).
    Corresponding colorbars presented in Figure~\ref{fig10:}(a). 
    
\end{itemize}

% % %%%%%
% % % APPENDIX C - SUPPLEMENTARY MATERIAL DESCRIPTION
% % %%%%
% \clearpage
% \input{sections/AppC-WaveCharacteristics}

% % %%%%%
% % % APPENDIX D - IW BEAM ANGLE
% % %%%%
% \clearpage
% \input{sections/AppD-IWBeamAngle}

\end{document}